\PassOptionsToPackage{dvipsnames,table}{xcolor}

\documentclass[acmsmall,screen,nonacm]{acmart}

\usepackage{amsmath}
\usepackage{hyperref}
\usepackage{mathpartir}
\usepackage{epigraph}
\usepackage{xspace}
\usepackage{tikz-cd}
\usepackage{tikzit}

\tikzstyle{square}=[fill={rgb,255: red,205; green,255; blue,205}, draw=black, shape=rectangle]
\tikzstyle{circle}=[fill=white, draw=black, shape=circle]
\tikzstyle{grey_square}=[fill={rgb,255: red,229; green,229; blue,229}, draw=black, shape=rectangle]
\tikzstyle{red_square}=[fill={rgb,255: red,255; green,205; blue,205}, draw=black, shape=rectangle]
\tikzstyle{big box}=[fill=white, draw=black, shape=rectangle, minimum width=8.5cm, minimum height=4cm]
\tikzstyle{light grey square}=[fill={rgb,255: red,245; green,245; blue,245}, draw=black, shape=rectangle]
\tikzstyle{green_circle}=[fill={rgb,255: red,205; green,255; blue,205}, draw=black, shape=circle]
\tikzstyle{red_circle}=[fill={rgb,255: red,255; green,205; blue,205}, draw=black, shape=circle]

\tikzstyle{arrow}=[->, fill=none]
\tikzstyle{reverse arrow}=[<-]

\usepackage{float}
\usepackage{listings}
\lstdefinelanguage{myC}{
  language=C,
  basicstyle=\linespread{1.2}\ttfamily,
  showspaces=false,              
  showstringspaces=false,        
  showtabs=false,    
  tabsize=2,                      
  captionpos=b,                   
  breaklines=true,                
  breakatwhitespace=false,
  escapeinside={\%*}{*)},        
  keywordstyle=\bfseries\color{black},    
  numberstyle=\tiny\color{gray},
}
\lstdefinelanguage{myinlineC}{
  language=myC,
  basicstyle=\ttfamily
}
\lstdefinelanguage[x86gasm]{Assembler}[x86masm]{Assembler}{%
,basicstyle=\ttfamily\singlespacing
,morekeywords={rax,rbx,rcx,rdx,rip,rdi,rsi,rsp,subq,decl,movq
              ,movl,xorl,imull,popq,popl,pushl}%
,morekeywords=[2]{.file,.section,.string,.text,.globl,.cfi_startproc
                 ,.cfi_def_cfa_offset,.cfi_endproc,.size,.ident}%
}
\definecolor{cmmtcolor}{named}{OliveGreen}

\lstdefinelanguage{Coq}{
,morekeywords={match,end,Definition,Inductive,Lemma,Theorem,Record,
               Hypothesis,Variable,Section,End,case,of,if,then,else,
               is,let,in,do,return,with,Extract,Constant,Inlined,Inline,
               Extraction,Fixpoint,Program,Function,Class,CoInductive,
               CoFixpoint,Variant,Instance,Context}
,morecomment=[s]{(*}{*)}
,keywordstyle=\bfseries\color{MidnightBlue}
,commentstyle={\color{cmmtcolor}}
,basicstyle=\small\linespread{1.0}\sffamily
,columns=fullflexible
,numberstyle=\tiny\color{gray}
,escapeinside={@}{@}
,literate=
    {:=}{{$\triangleq\;$}}1
    {<-}{{$\leftarrow\;$}}1
    {=>}{{$\Rightarrow\;$}}1
    {->}{{$\rightarrow\;$}}1
    {<->}{{$\leftrightarrow\;$}}1
    {<==}{{$\leq\;$}}1
    {\\/}{{$\vee\;$}}1
    {/\\}{{$\land\;$}}1
    {ffun}{{$\mathsf{ffun}$}}1    
    {fun}{{$\lambda$}}1
    {forall}{{$\forall$}}1
    {exists}{{$\exists$}}1
    {Z}{{$\mathbb{Z}$}}1
    {Z0}{{$\mathbb{Z}_0$}}1
    {<=}{{$\leq\;$}}1
    {>=}{{$\geq\;$}}1
    {<>}{{$\neq\;$}}1                
}

\lstdefinelanguage{PPL}{
,morekeywords={if,then,else,return,observe,while}
,keywordstyle=\bfseries\color{MidnightBlue}
,commentstyle={\color{cmmtcolor}}
,basicstyle=\linespread{1.2}\sffamily
,columns=fullflexible
,numberstyle=\tiny\color{gray}
,escapeinside={@}{@}
,literate=
    {<-}{{$\leftarrow\;$}}1
}

\lstdefinelanguage{cpGCL}{
,morekeywords={if,then,else,observe,while,do,end,skip,uniform}
,keywordstyle=\bfseries\color{MidnightBlue}
,morecomment=[s]{(*}{*)}
,commentstyle={\color{cmmtcolor}}
,basicstyle=\linespread{1.0}\sffamily
,columns=fullflexible
,numberstyle=\tiny\color{gray}
,escapeinside={@}{@}
,literate=
    {<-}{{$\leftarrow\;$}}1
}

\lstdefinelanguage{grammar}{
,morekeywords={skip,abort,observe,ite,while}
,keywordstyle=\bfseries\color{MidnightBlue}
,commentstyle={\color{cmmtcolor}}
,basicstyle=\linespread{1.0}\ttfamily
,numberstyle=\tiny\color{gray}
,escapeinside={@}{@}
}

\newcommand{\cwpcompact}{34}
\newcommand{\cwpco}{122}

\newcommand{\cwpcozexistszunique}{626}
\newcommand{\cwpcoopzexistszunique}{758}
\newcommand{\cwpcozincl}{197}

\newcommand{\cwpcontinuouszco}{300}

\newcommand{\cwpcozunique}{375}

\newcommand{\cwpcoopzintro}{852}

\newcommand{\cwpcozco}{1105}

\newcommand{\cwpcozcoopP}{1229}
\newcommand{\cwpProperzco}{800}

\newcommand{\cwpcontinuouszcozincl}{971}
\newcommand{\cwpcontinuouszind}{997}

\newcommand{\cwpDense}{45}
\newcommand{\cwpaCPO}{51}
\newcommand{\cwpaCPOzprod}{85}
\newcommand{\cwpaCPOzsum}{115}
\newcommand{\cwpcoiter}{827}
\newcommand{\cwpconatztozconatx}{111}
\newcommand{\cwpconatxztozconat}{117}
\newcommand{\cwpconatzext}{205}
\newcommand{\cwpconatzfinitezorzomega}{785}
\newcommand{\cwpCompactznat}{738}
\newcommand{\cwpaCPOzconat}{751}
\newcommand{\cwpnatzinj}{97}
\newcommand{\cwpconatzprefix}{240}
\newcommand{\cwpnatziter}{793}
\newcommand{\cwpconat}{34}
\newcommand{\cwpconatzeq}{123}
\newcommand{\cwpomega}{38}
\newcommand{\cwpconatzeqzaxiom}{202}
\newcommand{\cwpcozero}{35}
\newcommand{\cwpcosucc}{36}
\newcommand{\cwpconatzeqzzero}{124}
\newcommand{\cwpconatzeqzsucc}{125}
\newcommand{\cwpcofold}{1212}

\newcommand{\cwpafilter}{1530}
\newcommand{\cwpcofilter}{1534}

\newcommand{\cwpcolistzexists}{862}

\newcommand{\cwpcolistzforall}{868}
\newcommand{\cwplistzcolistzle}{1031}
\newcommand{\cwpcolistzlex}{1049}

\newcommand{\cwpcolistzlength}{1209}
\newcommand{\cwpproductivexx}{1273}

\newcommand{\cwpbadzbool}{1844}

\newcommand{\cwpcosumzcons}{1784}
\newcommand{\cwpcolistzexistszintrow}{884}
\newcommand{\cwpcolistzexistszintroe}{894}
\newcommand{\cwpcolistzexistszelim}{911}

\newcommand{\cwpcolistzforallzintro}{928}
\newcommand{\cwpcolistzforallzelimw}{942}
\newcommand{\cwpcolistzforallzelime}{951}
\newcommand{\cwpcolistzlezcolistzlex}{1089}
\newcommand{\cwpcofilterzcons}{1542}
\newcommand{\cwpcofilterzcomm}{1563}
\newcommand{\cwpcofoldzcons}{1222}

\newcommand{\cwpbadzstreamzproductive}{1848}
\newcommand{\cwpbadzstreamzspec}{1862}
\newcommand{\cwplistzforallzforallzafilter}{1684}
\newcommand{\cwpcolistzforallzcofilterx}{1695}

\newcommand{\cwpCompactzlist}{665}

\newcommand{\cwpaCPOzcolist}{678}
\newcommand{\cwpmonotonezfold}{712}
\newcommand{\cwpantimonotonezfold}{742}
\newcommand{\cwplistzle}{169}

\newcommand{\cwpfold}{693}
\newcommand{\cwpcolist}{34}
\newcommand{\cwpcolistzle}{47}
\newcommand{\cwpbadzstream}{1846}

\newcommand{\cwpconil}{35}
\newcommand{\cwpcocons}{36}
\newcommand{\cwplistzleznil}{170}
\newcommand{\cwplistzlezcons}{171}
\newcommand{\cwpcolistzleznil}{48}
\newcommand{\cwpcolistzlezcons}{49}

\newcommand{\cwpcotreeziter}{1396}
\newcommand{\cwpcotreezbind}{1211}

\newcommand{\cwptcofold}{1074}

\newcommand{\cwpatreezcotreezfilter}{1333}
\newcommand{\cwpcotreezfilter}{1336}

\newcommand{\cwptcofoldzleaf}{1083}

\newcommand{\cwpcotreezfilterzleaf}{1580}

\newcommand{\cwpcotreeziterzunfold}{1520}
\newcommand{\cwpCompactzatree}{829}

\newcommand{\cwpaCPOzcotree}{842}

\newcommand{\cwpatree}{74}

\newcommand{\cwptfold}{929}
\newcommand{\cwpcotree}{48}
\newcommand{\cwpcotreezle}{170}

\newcommand{\cwpcobot}{49}
\newcommand{\cwpcoleaf}{50}
\newcommand{\cwpconode}{51}
\newcommand{\cwpabot}{75}
\newcommand{\cwpaleaf}{76}
\newcommand{\cwpanode}{77}
\newcommand{\cwpcotreezlezbot}{171}
\newcommand{\cwpcotreezlezleaf}{172}
\newcommand{\cwpcotreezleznode}{173}

\newcommand{\cwpdirected}{115}
\newcommand{\cwpdownwardzdirected}{119}
\newcommand{\cwpchain}{104}
\newcommand{\cwpmonotone}{320}
\newcommand{\cwpantimonotone}{325}
\newcommand{\cwpcontinuous}{545}
\newcommand{\cwpcocontinuous}{554}
\newcommand{\cwpequ}{238}
\newcommand{\cwpdeczcontinuous}{588}
\newcommand{\cwpdeczcocontinuous}{595}

\newcommand{\cwpOType}{28}
\newcommand{\cwpPType}{62}

\newcommand{\cwpPTypezProp}{128}
\newcommand{\cwpTTypezProp}{134}

\newcommand{\cwpsup}{93}

\newcommand{\cwpinf}{106}

\newcommand{\cwpCPO}{26}
\newcommand{\cwplCPO}{31}

\newcommand{\cwpLatticezProp}{156}
\newcommand{\cwpLatticezbool}{207}

\newcommand{\cwppLatticezeR}{379}

\newcommand{\cwpasievezaux}{41}
\newcommand{\cwpsievezaux}{49}
\newcommand{\cwpsievezcons}{441}
\newcommand{\cwpsievezcomplete}{141}
\newcommand{\cwpsievezsound}{186}
\newcommand{\cwpsortedzsieve}{410}
\newcommand{\cwpnodupzsieve}{418}
\newcommand{\cwpproductivezsieve}{289}
\newcommand{\cwpnats}{36}
\newcommand{\cwpexzprimezgt}{601}
\newcommand{\cwpbtwp}{138}
\newcommand{\cwpcotwp}{145}

\newcommand{\cwpcotwpzleaf}{196}

\newcommand{\cwpcotwpzbind}{216}

\newcommand{\cwpmarkovzinequality}{593}
\newcommand{\cwpasum}{143}
\newcommand{\cwptcosum}{146}
\newcommand{\cwpatreezlang}{177}
\newcommand{\cwpcotreezlang}{180}
\newcommand{\cwpcotreezpreimage}{183}
\newcommand{\cwptcosumzleaf}{162}
\newcommand{\cwpcotreezlangzleaf}{196}
\newcommand{\cwpcotwpzmuzlang}{288}
\newcommand{\cwpcotwpztcosumzpreimage}{348}

\newcommand{\cwpmu}{399}
\newcommand{\cwpuniform}{402}
\newcommand{\cwpproduceszinzsigmaqw}{423}
\newcommand{\cwpcotreezsampleszequidistributed}{513}
\newcommand{\cwpconcat}{1438}
\newcommand{\cwptconcat}{1341}
\newcommand{\cwpstar}{1547}

\newcommand{\cwpepsilon}{1251}
\newcommand{\cwpinzlangzcotriezle}{1215}
\newcommand{\cwpcotriezlezorder}{2673}
\newcommand{\cwpCompactztrie}{552}
\newcommand{\cwpaCPOzcotrie}{911}
\newcommand{\cwpKleeneAlgebraLawszlang}{2987}
\newcommand{\cwptrie}{38}
\newcommand{\cwptriezle}{61}
\newcommand{\cwpiszbot}{42}
\newcommand{\cwpcotrie}{174}
\newcommand{\cwpcotriezle}{197}
\newcommand{\cwpcotriezbot}{222}
\newcommand{\cwpunion}{1295}
\newcommand{\cwpintersection}{1307}
\newcommand{\cwpcomplement}{1323}
\newcommand{\cwpcotrieznode}{175}
\newcommand{\cwpcotriezleznode}{198}
\newcommand{\cwptriezbot}{39}
\newcommand{\cwptrieznode}{40}
\newcommand{\cwptriezlezbot}{62}
\newcommand{\cwptriezleznode}{63}
\newcommand{\cwpiszbotzbot}{43}
\newcommand{\cwpiszbotznode}{44}

\theoremstyle{plain}
\newtheorem{theorem}{Theorem} 
\theoremstyle{definition}
\newtheorem{definition}[theorem]{Definition} 

\newtheorem{lemma}[theorem]{Lemma}
\newtheorem{corollary}[theorem]{Corollary}
\newtheorem{example}[theorem]{Example}
\newtheorem{axiom}{Axiom}
\theoremstyle{remark}
\newtheorem{remark}{Remark}

\newcommand{\link}[3]{\href{https://github.com/bagnalla/algco/blob/main/#1.v\#L#2}{#3}}

\newcommand{\R}{\mathbb{R}}
\newcommand{\Rpos}{\mathbb{R}_{\geq 0}}
\newcommand{\eR}{\mathbb{R}^\infty_{\geq 0}}

\newcommand{\alist}[1]{\mathcal L_{#1}}
\newcommand{\colist}[1]{\mathcal L^{*}_{#1}}
\newcommand{\atree}[1]{\mathcal T_{#1}}
\newcommand{\cotree}[1]{\mathcal T^{*}_{#1}}
\newcommand{\tlang}{\mathsf{tlang}_\Sigma}
\newcommand{\lang}{\mathsf{lang}_\Sigma}
\newcommand{\bind}{\gg\!\!=}

\newcommand{\co}[1]{#1^{\mathsf{co}}}
\newcommand{\coop}[1]{#1^{\mathsf{\hat{co}}}}
\newcommand{\basis}[1]{\mathsf{B}(#1)}
\newcommand{\cons}[1]{\mathsf{\mathbf{#1}}}

\newcommand{\prop}{\mathbb{P}}
\newcommand{\pred}[1]{#1 \rightarrow \prop}
\newcommand{\rel}[2]{#1 \rightarrow #2 \rightarrow \prop}
\newcommand{\bool}{\mathbb{B}}
\newcommand{\nat}{\mathbb{N}}
\newcommand{\conat}{\co{\nat}}

\newcommand{\true}{\mathbf{true}}
\newcommand{\false}{\mathbf{false}}

\newcommand{\keyw}[1]{\ensuremath{\mathsf{#1}\xspace}}
\newcommand{\zar}{\keyw{Zar}\xspace}

\begin{document}


\title{Inductive Reasoning for Coinductive Types}


\author{Alexander Bagnall}
\email{abagnalla@gmail.com}
\affiliation{
  \institution{Ohio University}
  \city{Athens}
  \state{Ohio}
  \country{USA}}
\orcid{0000-0001-6593-0661}
\author{Gordon Stewart}
\affiliation{
  \institution{BedRock Systems, Inc.}
  \city{Boston}
  \state{Massachusetts}
  \country{USA}}
\email{gordon@bedrocksystems.com}
\orcid{0000-0003-0244-2980}
\author{Anindya Banerjee}
\affiliation{
  \institution{IMDEA Software Institute}
  \city{Pozuelo de Alarcon}
  \state{Madrid}
  \country{Spain}}
\email{anindya.banerjee@imdea.org}
\orcid{0000-0001-9979-1292}

\setcopyright{none}
\authorsaddresses{} 

\thispagestyle{plain}

\begin{abstract}
  We present AlgCo (Algebraic Coinductives), a practical framework for
  inductive reasoning over commonly used coinductive types such as
  conats, streams, and infinitary trees with finite branching
  factor. The key idea is to exploit the notion of \textit{algebraic
  complete partial order} from domain theory to define continuous
  operations over coinductive types via primitive recursion on
  ``dense'' collections of their elements, enabling a convenient
  strategy for reasoning about algebraic coinductives by
  straightforward proofs by induction. We implement the AlgCo
  framework in Coq and demonstrate its power by verifying a stream
  variant of the sieve of Eratosthenes, a regular expression library
  based on coinductive trie encodings of formal languages, and
  expected value semantics for coinductive sampling processes over
  discrete probability distributions in the random bit model.
\end{abstract}

\settopmatter{printacmref=false}
\setcopyright{none}
\renewcommand\footnotetextcopyrightpermission[1]{}
\pagestyle{plain}
\maketitle


%

\section{Introduction}
\label{sec:intro}


As a tool for defining and proving correctness of computations over
well-founded data, the principle of induction is intimately familiar
to most computer scientists. Consequently, proof assistants like Coq
and Agda provide ergonomic support for programming and proving with
induction~\cite{inria2022coq,agda2022doc}. The dual principle
of \textit{coinduction}~\cite{kozen2017practical}, on the other hand,
is not nearly as well supported.

Coinduction provides a natural means for programming with and
verifying properties of infinitary structures such as conats (natural
numbers extended with a ``point at infinity''), streams (infinite
lists)~\cite{chlipala2022certified}, and potentially nonterminating
decision processes (e.g., samplers for discrete distributions compiled
from probabilistic
programs~\cite{https://doi.org/10.48550/arxiv.2211.06747}). However,
coinductive reasoning has a tendency to betray intuition, and proof
assistants like Coq and Agda are designed with a noticeable bias
toward induction, which can exacerbate the inherently unintuitive
nature of coinduction.  As a result, the use of coinduction in a proof
assistant is often plagued by technical snags due to rigid syntactic
guardedness conditions (limiting the range of allowable coinductive
definitions), and the generation of coinduction hypotheses that
interact poorly with automation
tactics~\cite{DBLP:conf/popl/HurNDV13,chlipala2022certified}). Hur et
al. proposed -- and Pous later
generalized~\cite{DBLP:conf/lics/Pous16} -- parameterized coinduction
($\mathsf{paco}$~\cite{DBLP:conf/popl/HurNDV13}) as a way to address
rigid syntactic checks through the use of a semantic notion of
guardedness.  While $\mathsf{paco}$ substantially upgrades coinduction
in Coq, it does not represent a fundamental departure from primitive
coinduction.

This paper proposes an alternative strategy: instead of generalizing
coinduction in Coq (e.g., as $\mathsf{paco}$ does to semantic
guardedness), we consider a subset of coinductive types -- those
corresponding to algebraic
CPOs~\cite{DBLP:journals/tcs/Jung90,gunter1992semantics} -- for which
a completely different strategy can be applied for reasoning about
coinductive programs that is more suited to the inherently inductive
disposition of proof assistants like Coq.

To illustrate our approach, consider the problem of defining a
$\mathsf{filter}$ operation for coinductive streams (i.e., infinite
lists). In Haskell, it is possible to filter a stream by a given
predicate -- e.g., the stream of even numbers is given by $[n \: | \:
n \leftarrow [0..], n \: `mod` \: 2 = 0]$. We can attempt to implement
a similar $\mathsf{filter}$ operation for streams in Coq as follows:
\begin{center}
  \begin{tabular}{c}
    \begin{lstlisting}[language=coq,mathescape=true]
CoFixpoint filter (P : $\nat$ -> $\bool$) (s : Stream $\nat$) :=
  match s with Cons n s' =>
    if P n then Cons n (filter P s') else filter P s'
  end.
    \end{lstlisting}
  \end{tabular}
\end{center}


This definition is rejected, however, due to presence of an unguarded
recursive call (not wrapped in a $\cons{Cons}{}$ constructor) in the
`else' branch. Indeed, it is a good thing that it is rejected, or else
we could create a divergent term (rendering the logic of Coq
inconsistent) by filtering any stream by $P
:= \lambda\,\_. \: \false$.  While there may be specific circumstances
in which we could prove it safe to filter a stream by a given
predicate (i.e., when the resulting stream will be
``productive''~\cite{chlipala2022certified}), we cannot do that
because $\mathsf{filter}$ is not definable in the first place.




A common solution (taken, e.g., by Xia et al. in the interaction
tree library~\cite{xia2019interaction}) is to add a constructor to the
stream type for so-called ``silent steps'' (\textit{Tau} nodes). Tau
nodes trivially satisfy the guardedness checker -- wrap unguarded
co-recursive calls by applications of Tau -- but lead to 
extra cases in definitions and proofs, and to unnecessary execution overhead.
Moreover, replacing points of divergence with 
infinite sequences of Taus passes the responsibility of handling
divergence to consumers of the stream, leading to complications in
subsequent computation and analysis.

As an example of a complication of Tau,
consider taking the infinite sum of a stream of reals in which Tau
nodes can appear. Lacking a general induction principle for streams,
we resort to a coinductive relation between streams and their sums
(where $\eR$ denotes the nonnegative reals extended with $+\infty$):


\begin{center}
  \begin{tabular}{c}
    \begin{lstlisting}[language=coq,mathescape=true]
CoInductive sum : Stream $\eR$ -> $\eR$ -> $\prop$ :=
| sum_tau : forall $ $ s r, sum s r -> sum (Tau s) r
| sum_cons : forall $ $ s x r, sum s r -> sum (Cons x s) (x + r).
    \end{lstlisting}
  \end{tabular}
\end{center}

\noindent
and attempt to prove that the relation is functional:
\begin{center}
  \begin{tabular}{c}
    \begin{lstlisting}[language=coq,mathescape=true]
Lemma sum_@{fun}@ctional (s : Stream $\eR$) (a b : $\eR$) :
  sum s a -> sum s b -> a = b.
Proof. ... Abort. (* unprovable *)
    \end{lstlisting}
  \end{tabular}
\end{center}


\noindent
But $\mathsf{sum\_functional}$ is unprovable because it isn't true.
The problem is that coinductive relations are interpreted as the
\textit{greatest} relations closed under their rules, and so tend to
relate more pairs of elements than intended. E.g., the stream $\Omega
\triangleq \mathsf{Tau}\,\: \Omega$ is related by $\mathsf{sum}$ to
every $r : \R$, so sum is clearly not a functional (i.e.,
deterministic) relation because, e.g., $\mathsf{sum} \: \Omega \: 0$
and $\mathsf{sum} \: \Omega \: 1$, but $0 \neq 1$). We could constrain
the lemma to apply only to streams containing no infinite sequences of
Taus, but then we take on the burden of proving this side condition
for all of our stream constructions. Defining the $\mathsf{sum}$
relation via $\mathsf{paco}$ does not help because it suffers from the
same fundamental problem, being defined as the
\textit{greatest} fixed point of a monotone functional.



We take an alternative approach (similar to that
in~\cite{DBLP:conf/ecoop/RusuN22}) for defining $\mathsf{sum}$
inspired by domain theory. We first define an inductive analogue of
$\mathsf{sum}$ on lists:
\begin{center}
  \begin{tabular}{c}
    \begin{lstlisting}[language=coq,mathescape=true]
Fixpoint lsum (l : List $\eR$) : $\eR$ :=
  match l with
  | nil => 0
  | cons x l' => x + lsum l'
  end.
    \end{lstlisting}
  \end{tabular}
\end{center}

\noindent
and then for $s : \mathsf{Stream} \: \eR$ we define $\mathsf{sum} \: s
: \eR \triangleq \mathsf{sup} \: \{ \mathsf{lsum} \: l \mid l \text{
is a finite prefix of } s \}$. This definition of $\mathsf{sum}$ maps
every stream to a unique element of $\eR$, even in the presence of Tau
nodes. We may ask: under what conditions is it possible to define
functional mappings on coinductives in this way? For the answer we
turn to a fundamental result of domain theory
(Lemma~\ref{lemma:continuous-extension}):
\begin{itemize}
  \item[-] $\eR$ is a CPO (complete partial order),
  \item[-] $\mathsf{Stream \: \eR}$ is an \textit{algebraic} CPO,
  \item[-] and $\mathsf{nsum}$ is \textit{monotone}.
\end{itemize}

We call $\mathsf{sum}$ the \textit{continuous extension} of
$\mathsf{lsum}$. It is continuous by construction, and many proofs
about it (specifically, proofs of \textit{continuous properties}, see
Section~\ref{subsec:cocontinuous-properties}) can be reduced to
straightforward proofs by induction over
lists. Moreover, \textit{every continuous function} from streams into
$\eR$ (or indeed, from any algebraic CPO into another CPO) can be
obtained in this way, that is, as the continuous extension of a
monotone function over finite elements.

A basic connection between coinductive types and CPOs that can be
understood as follows: A coinductive type, interpreted as the final
coalgebra of a given functor~\cite{DBLP:journals/jsc/Hagino89}, is
equipped (by finality) with a canonical introduction principle for
defining elements of the type. The completeness property of a CPO can
likewise be viewed as a kind of introduction principle for which
elements of the CPO are introduced as suprema of directed sets of
approximations. Thus coinductive types naturally form CPOs (e.g., the
type of streams can be seen as the ``completion'' of the type of lists
wrt.~suprema of directed
sets~\cite{DBLP:journals/lmcs/Adamek21}). Furthermore, by exploiting
algebraicity (the existence of a dense compact subset) of the domain
and completeness of the codomain (in this case
$\mathsf{Stream} \: \eR$ and $\eR$, respectively), we are able to
provide a \textit{continuous elimination principle} for algebraic
domains (Lemma~\ref{lemma:continuous-extension}) for defining
continuous functions like $\mathsf{filter}$ and $\mathsf{sum}$ as well
as continuous predicates and relations
(Section~\ref{subsec:cocontinuous-properties}) over them.

While the theory of algebraic CPOs is well known to domain
theorists~\cite{gunter1992semantics}, the practical implications of
its connection to coinductive types appear under-appreciated. We are
not aware of any prior work to explicitly draw this connection and
elaborate on its practical applications in a theorem proving
environment. The main contribution of the present paper is thus to
elucidate the power and generality of the concept of algebraic CPOs by
providing a comprehensive framework (\textit{AlgCo}, short
for \textit{Algebraic Coinductives}) for programming and proving with
continuous functions in Coq over the class of coinductive types
forming algebraic CPOs (including commonly used structures such as
conats, streams, and infinitary trees), and demonstrating its utility
on a handful of interesting use cases including the semantics of
probabilistic programming languages (Section~\ref{sec:cotrees}). In
Section~\ref{subsec:cofolds}, we show that the aforementioned
$\mathsf{filter}$ operation can be defined with the tools of AlgCo
without the need for Tau nodes at all.

\subsection{Contributions}

This paper makes the following contributions:

\smallskip
\textbf{Concept.}
We describe AlgCo (short for Algebraic Coinductives): a comprehensive
type-theoretic framework for programming with continuous functions
over a class of structures known as algebraic CPOs. AlgCo allows
definitions and proofs over algebraic CPOs to be reduced to primitive
recursion and proofs by induction over finite approximations. We
provide an implementation of the theory in Coq called
\href{https://github.com/bagnalla/algco}{AlgCo.}

\smallskip
\textbf{Technical.}
The technical contributions of this paper include:
\begin{itemize}

  \item[-] an introduction to the basic concepts and proof principles
  of algebraic coinductives (Section~\ref{sec:algebraic-coinductives}),

  \item[-] a framework for \textit{lazy coiteration} with conats
  (Section~\ref{sec:conats}),

  \item[-] a library for reasoning about coinductive streams, with
  application to a formally verified sieve of Eratosthenes
  (Section~\ref{sec:colists}),

  \item[-] verified regular expression matchers based on coinductive
  tries (Section~\ref{sec:cotries}), and

  \item[-] a library for reasoning about coinductive binary trees,
  with application to (real-valued) expected value semantics of
  potentially nonterminating sampling processes in the random bit
  model (Section~\ref{sec:cotrees}), culminating in a theorem showing
  equidistribution of generated samples wrt.~that semantics (Theorem
  ~\ref{theorem:equidistribution}).
\end{itemize}

{\textbf{Source Code.}} Embedded hyperlinks in the PDF point to the
underlying \href{https://github.com/bagnalla/algco}{Coq sources.}


\subsection{Limitations}

Continuous extensions such as $\mathsf{sum}
: \mathsf{Stream} \: \eR \rightarrow \eR$ are defined
via a nonconstructive supremum operator (see
Section~\ref{subsec:domain-theory}) and thus are not computable in
general. In some special cases, however, including lazy coiteration
(Section~\ref{subsec:infinite-fuel}) and cofolds
(Section~\ref{subsec:cofolds})), we provide Haskell extraction primitives that
are useful in practice.
Section~\ref{subsec:extracting-sieve} discusses
computability in more detail.

The techniques described in this paper apply only to types forming
algebraic CPOs. While this includes many useful coinductive types such
as streams and infinitary binary trees, there exist many interesting
types for which it does not apply, e.g., trees that are infinite in
both depth $\textit{and breadth}$.







{\textbf{Axiomatic Base.} We extend the type theory of Coq with
excluded middle, constructive indefinite description, and functional
extensionality~\cite{chargueraud2017axioms,DBLP:conf/itp/Chargueraud10}. We
also use Coq's axiomatic real number library and extensionality axioms
(e.g., Axiom~\ref{axiom:conat-extensionality} in
Section~\ref{sec:conats}) for coinductive types (see
Appendix~\ref{app:coinductive-extensionality}).

\section{Algebraic Coinductives}
\label{sec:algebraic-coinductives}

AlgCo is a formal framework for defining and reasoning about
Scott-continuous functions on a class of structures known as algebraic
CPOs (Section~\ref{subsec:algebraic-cpos}). Examples of algebraic CPOs
include conats (Example~\ref{def:conat}), streams
~\cite{chlipala2022certified,kozen2017practical}
(Section~\ref{sec:colists}), countable sets, and infinitary trees with
fixed and finite branching factor (Section~\ref{sec:cotrees}).

Much of the basic theory underlying the framework of algebraic
coinductives is well-known and can be studied in classic texts on
domain theory~\cite{gunter1992semantics,abramsky1994domain}. However,
some of its practical implications for formal verification in type
theory remain unexplored. In this section, we show how a fundamental
result about algebraic CPOs (Lemma~\ref{lemma:continuous-extension})
gives rise to a powerful framework for programming with coinductive
types in Coq.

\subsection{Domain Theory}
\label{subsec:domain-theory}

We begin with a quick tour of some definitions from order and domain
theory upon which the rest of the paper depends.

\begin{definition}[\link{order}{\cwpOType}{Ordered type}]
  \label{def:ordered-type}
  
  An \textit{ordered type} (or \textit{partial order}) is a type $A$
  with an order relation $\sqsubseteq_A$ that is reflexive and
  transitive.
\end{definition}


We often view $\sqsubseteq_A$ as an approximation relation, such that
$a \sqsubseteq b$ when $a$ is a somehow coarser or less defined
version of $b$. An ordered type $A$ is said to be \textit{pointed}
when it contains a bottom element $\bot_A$ such that $\forall \: a :
A, \bot_A \sqsubseteq_A a$. $A$ may also have a top element $\top_A$
such that $\forall \: a : A, a \sqsubseteq_A \top_A$. We often omit
the subscripts on $\sqsubseteq_A$ and $\bot_A$ when the ordered type
$A$ is clear from context.

\begin{definition}[\link{order}{\cwpequ}{Order equivalence}]
  \label{def:order-equivalence}

  Elements $x$ and $y$ of ordered type $A$
  are \textit{order-equivalent}, written $x \simeq y$, when
  $x \sqsubseteq y$ and $y \sqsubseteq x$.
\end{definition}

  

\begin{definition}[\link{order}{\cwpdirected}{Directed set}]
  \label{def:directed-set}
  
  For index type $I$ and ordered type $A$, a collection $U :
  I \rightarrow A$ is \textit{directed} when $\forall \: i \: j :
  I, \exists \: k : I, U \: i \sqsubseteq U \: k \land U \:
  j \sqsubseteq U \: k$,
  or \link{order}{\cwpdownwardzdirected}{\textit{downward-directed}}
  when $\forall \: i \: j : I, \exists \: k : I, U \: k \sqsubseteq
  U \: i \land U \: k \sqsubseteq U \: j$.
\end{definition}

Intuitively, a directed set is one in which all elements are
ultimately approximating the same thing. This leads to a natural
notion of convergence: a directed set ``converges'' to its supremum
whenever it exists. A special case of directed set is that of
$\omega$-chain:

\begin{definition}[\link{order}{\cwpchain}{$\omega$-chain}]
  \label{def:omega-chain}

  For ordered type $A$, a collection $C : \nat \rightarrow A$ is an
  $\omega$-chain when $\forall \: i \: j : \nat, i \le j \Rightarrow
  C \: i \sqsubseteq C \: j$.
\end{definition}

\begin{definition}[\link{cpo}{\cwpCPO}{Complete partial order}]
  \label{def:cpo}
  
  An ordered type $A$ is a \textit{complete partial order} (CPO) when
  every directed $U : \nat \rightarrow A$ has a supremum, or
  a \link{cpo}{\cwplCPO}{\textit{lower-complete partial order}}
  (lCPO) when every downward-directed $V : \nat \rightarrow A$ has an
  infimum. We say that $A$ is a \textit{d-lattice} when it is both a
  CPO and lCPO.
\end{definition}

Our choice to specialize to countable directed sets is an artifact of
the formalization. We could work with $\omega$-CPOs instead (where
every $\omega$-chain has a supremum) but it is generally easier to
construct directed sets than $\omega$-chains (e.g., every collection
of real numbers is trivially directed).

\begin{remark}[\link{cpo}{\cwpLatticezProp}{$\prop$ is a d-lattice}]
  \label{remark:prop-lattice}

  The type $\prop$ of propositions ordered by implication (i.e.,
  $P \sqsubseteq Q \iff P \Rightarrow Q$) is a d-lattice
  with \link{order}{\cwpPTypezProp}{bottom $\bot$}
  and \link{order}{\cwpTTypezProp}{top $\top$} where for any $U :
  I \rightarrow \prop$ (not necessarily (downward-)directed),
  $\sup{(U)} \triangleq \exists \: i. U \: i$ and
  $\inf{(U)} \triangleq \forall \: i. U \: i$.
\end{remark}

\begin{remark}[\link{cpo}{\cwpLatticezbool}{$\bool$ is a d-lattice}]
  \label{remark:bool-lattice}

  The type $\bool$ of Booleans is a d-lattice with bottom $\false$ and
  top $\true$.
\end{remark}

\begin{remark}[\link{eR}{\cwppLatticezeR}{$\eR$ is a d-lattice}]
  \label{remark:eR-lattice}

  The type $\eR$ of nonnegative extended reals is a d-lattice with
  bottom $0$ and top $+\infty$.
\end{remark}

\begin{definition}[\link{order}{\cwpmonotone}{Monotone}]
  \label{def:monotone}

  For ordered types $A$ and $B$, a function $f : A \rightarrow B$
  is \textit{monotone} when $\forall \: x \: y : A, x \sqsubseteq
  y \Rightarrow f \: x \sqsubseteq f \: y$,
  or \link{order}{\cwpantimonotone}{\textit{antimonotone}} when
  $\forall \: x \: y : A, x \sqsubseteq y \Rightarrow f \:
  y \sqsubseteq f \: x$.
\end{definition}

\begin{definition}[\link{order}{\cwpcontinuous}{Continuous}]
  \label{def:continuous}

  For ordered types $A$ and $B$, a function $f : A \rightarrow B$
  is \textit{continuous} when for every directed set $U
  : \nat \rightarrow A$, $f \: (\sup{U}) = \sup{(f \circ U)}$,
  or \link{order}{\cwpcocontinuous}{\textit{cocontinuous}} when $f \:
  (\sup{U}) = \inf{(f \circ U)}$.
\end{definition}

Note that (co-)continuity implies (anti)monotonicity.

\begin{definition}[\link{order}{\cwpdeczcontinuous}{Lower-continuous}]
  \label{def:lower-continuous}

  For ordered types $A$ and $B$, a function $f : A \rightarrow B$
  is \textit{lower-continuous} (\textit{l-continuous}) when for every
  downward-directed set $U : \nat \rightarrow A$, $f \: (\inf{U})
  = \inf{(f \circ U)}$,
  or \link{order}{\cwpdeczcocontinuous}{\textit{l-cocontinuous}} when
  $f \: (\inf{U}) = \sup{(f \circ U)}$.
\end{definition}

\paragraph*{Computing Suprema}
Suprema of directed sets are not generally computable. However, we can
use the axioms of indefinite description and excluded
middle~\cite{chargueraud2017axioms} to define a
$\link{cpo}{\cwpsup}{\mathsf{sup}}$
($\link{cpo}{\cwpinf}{\mathsf{inf}}$) operator (special cases of
Hilbert's epsilon operator~\cite{coq2023epsilon}) allowing succinct
expression of suprema (infima) in a classical style within
computational contexts in Coq~\cite{DBLP:conf/itp/Chargueraud10}. We
define \link{cpo}{\cwpsup}{$\mathsf{sup} : (\nat \rightarrow
A) \rightarrow A$} such that $\mathsf{sup} \: f$ is the supremum of
any directed $f : \nat \rightarrow A$,
and \link{cpo}{\cwpinf}{$\mathsf{inf} : (\nat \rightarrow
A) \rightarrow A$} such that $\mathsf{inf} \: f$ is the infimum of any
downward-directed $f : \nat \rightarrow A$. Any attempt to compute
with these operators will become blocked~\cite{leroy2015evaluation},
but in some special cases, they can be implemented by extraction
primitives for lazy execution in Haskell (see
Section~\ref{subsec:extracting-sieve} for discussion).




\subsection{Algebraic CPOs}
\label{subsec:algebraic-cpos}

A CPO is traditionally said to be \textit{algebraic} when it contains
a subset of ``basis'' elements that can be used to approximate any
element of the domain~\cite{gunter1992semantics}. These basis elements
are required to be \textit{compact}:

\begin{definition}[\link{aCPO}{\cwpcompact}{Compact element}]
  \label{def:compact-element}

  Let $A$ be an ordered type. An element $x : A$ is \textit{compact}
  when for every directed collection $U : \nat \rightarrow A$ such
  that $\sup{(U)} = x$, we have $\exists \: i, U \: i = x$.
\end{definition}

A compact element $x : A$, while not necessarily finite, is finitely
approximable, i.e., any directed set ``converging'' to $x$ contains a
finite subset that also converges to $x$ (see
Section~\ref{sec:cotries} for an algebraic CPO whose basis elements
are compact but not finite). A type $A$ is compact when every $x : A$
is compact. Compactness is essential because it coincides with the
presence of an induction principle; a key idea of this paper is to
reduce reasoning about continuous functions over coinductive types to
inductive reasoning over basis elements (see, e.g.,
Lemmas~\ref{lemma:cofilter-comm}, \ref{lemma:cofilter-forall}, \ref{lemma:wp-filter}, \ref{lemma:wp-mu-lang}
and Theorems~\ref{theorem:sieve-complete}, \ref{theorem:sieve-sound}).








We adapt the traditional definition of algebraicity to a
type-theoretic framework by saying that a CPO $A$ is algebraic when
there exists a basis type $B$ that can be injected into $A$ (hence is
like a subset of $A$) and is ``dense'' in $A$, i.e., all elements of
$A$ can be obtained as suprema of directed collections of elements in
$B$ injected into $A$.



\begin{definition}[\link{aCPO}{\cwpDense}{Dense}]
  \label{def:dense}

  Let $A$ be a CPO and $B$ an ordered type. $B$ is \textit{dense} in
  $A$ when there exist continuous operations
  \begin{align*}
    & \mathsf{incl_{B,A}} : B \rightarrow A \\
    & \mathsf{idl_{B,A}} : A \rightarrow \mathbb{N} \rightarrow B
  \end{align*}
  such that for all $a : A$,
  \begin{align*}
    & \text{$\mathsf{idl_{B,A}} \: a$ is an $\omega$-chain, and}
    & \sup{(\mathsf{incl_{B,A}} \circ \mathsf{idl_{B,A}} \: a)} = a.
  \end{align*}
\end{definition}

The $\mathsf{idl}$ operator (read ``ideal'') applied to element $x :
A$ produces an $\omega$-chain of basis elements whose injections into
$A$ converge to $x$. Strictly speaking, we should only require
$\mathsf{idl}$ to produce directed sets, but we choose to constrain to
$\omega$-chains in hopes of developing a useful notion of general
computability of continuous extensions in the future. We omit the
subscripts on $\mathsf{idl}$ and $\mathsf{incl}$ when they are clear
from context.


\begin{definition}[\link{aCPO}{\cwpaCPO}{Algebraic CPO}]
  \label{def:algebraic-cpo}

  Let $A$ be a CPO and $B$ an ordered type. $A$ is
  an \textit{algebraic CPO} (aCPO) with \textit{basis} $B$ when $B$ is
  compact and dense in $A$.
\end{definition}







We let $\basis{A}$ denote the basis of algebraic CPO $A$. Algebraic
CPOs are closed under the formation
of \link{prod}{\cwpaCPOzprod}{products}
and \link{sum}{\cwpaCPOzsum}{sums.} The type $A \rightarrow C$ is an
algebraic CPO when $A$ is finite and $C$ is an algebraic CPO.








\subsection{Continuous Extensions}
\label{subsec:continuous-extensions}


A key idea of the AlgCo framework is to define continuous functions on
algebraic CPOs as \textit{continuous extensions} of simpler monotone
functions on basis elements, characterized by the following lemma
(based on~\cite[Lemma 5.24]{gunter1992semantics}):

\begin{lemma}[\link{aCPO}{\cwpcozexistszunique}{Continuous extension}]
  \label{lemma:continuous-extension}

  Let $A$ be an algebraic CPO, $C$ a CPO, and $f
  : \basis{A} \rightarrow C$ a monotone function. Then, there exists a
  unique continuous function $\link{aCPO}{\cwpco}{\co{f}} :
  A \rightarrow C$ (the \textit{continuous extension} of basis
  function $f$) such that $\co{f} \circ \mathsf{incl} = f$. I.e., the
  following diagram commutes:
  \[
  \begin{tikzcd}
    \basis{A} \arrow[rrrdd, "f"] \arrow[dd, "\mathsf{incl}"'] &  &  &   \\
    &  &  &   \\
    A \arrow[rrr, "! \co{f}", dashed]               &  &  & C
  \end{tikzcd}
  \]
\end{lemma}

The main aspects of Lemma~\ref{lemma:continuous-extension} are
two-fold:

\smallskip
\textbf{Existence.}
Any monotone function $f : \basis{A} \rightarrow C$ defined on the
basis of an algebraic CPO can be extended to a continuous function
$\co{f} : A \rightarrow C$ on the whole domain $A$.

\smallskip
\textbf{Uniqueness.}
The extension is \textit{unique}. Any continuous function completing
the diagram must be equal to $\co{f}$.

Existence of $\co{f}$ constitutes a \textit{continuous elimination
scheme} for algebraic CPOs: to define a continuous function on
algebraic CPO $A$, it suffices instead to define a monotone basis
function $f : \basis{A} \rightarrow C$ (typically by induction, (see
Sections~\ref{sec:colists} and~\ref{sec:cotrees})) and extend it to
$\co{f} : A \rightarrow C$.




Moreover, uniqueness of $\co{f}$ implies that
\textit{every} continuous function $g : A \rightarrow C$ can be
represented as a continuous extension $\co{f} : A \rightarrow C$ for
some monotone basis function $f : \basis{A} \rightarrow C$, namely $f
= g \circ
\mathsf{incl}$ (see Corollary~\ref{corollary:representation}).

Lemma~\ref{lemma:continuous-extension} tells us that the continuous
functions on algebraic CPOs are precisely the functions that can be
``easily'' defined (as extensions of basis functions), and entails a
collection of useful proof principles (e.g.,
Lemma~\ref{lemma:equivalence-continuous-extension},
Theorem~\ref{theorem:fusion}, and more in
Appendix~\ref{app:algebraic-cpo-proof-principles}) for reasoning about
them. For example, the following lemma is often useful for proving
equivalence of continuous extensions:





\begin{lemma}[\link{aCPO}{\cwpProperzco}{Equivalence of continuous extensions}]
  \label{lemma:equivalence-continuous-extension}
  
  Let $A$ be an algebraic CPO, $C$ any CPO, and $f
  : \basis{A} \rightarrow C$ and $g : \basis{A} \rightarrow C$
  monotone functions on the basis of $A$. Then,
  \[ f = g \Rightarrow \co{f} = \co{g}. \]
\end{lemma}

Lemma \ref{lemma:equivalence-continuous-extension} is extremely useful
in practice: to show $\co{f} \: a = \co{g} \: a $ for all $a : A$, it
suffices to show $f \: b = g \: b$ for all $b : \basis{A}$, which can
often be proved by straightforward induction.


  


Besides the fact that \link{aCPO}{\cwpcontinuouszco}{continuous
extensions are continuous} (which should not be understated since ad
hoc proofs of continuity are often difficult and time consuming), we
have a fusion law for continuous extensions:

\begin{theorem}[\link{aCPO}{\cwpcozco}{Fusion}]
  \label{theorem:fusion}

  Let $A$ be an algebraic CPO, $B$ and $C$ CPOs, $f
  : \basis{A} \rightarrow B$ a monotone function, and $g :
  B \rightarrow C$ a continuous function. Then, $g \circ \co{f}
  = \co{(g \circ f)}$.
\end{theorem}

\subsection{Cocontinuous Extensions}
\label{subsec:cocontinuous-extensions}

We can also obtain \textit{cocontinuous} functions over an algebraic
CPO as \link{aCPO}{\cwpcoopzexistszunique}{\textit{cocontinuous
extensions} of antimonotone basis functions}. Every proof principle
for continuous extensions has an analogue for cocontinuous extensions.




\subsection{(Co-)continuous Properties}
\label{subsec:cocontinuous-properties}

Continuous functions with codomain $\prop$ are
called \textit{continuous properties}. Continuous properties on an
algebraic CPO $A$ can be defined as continuous extensions
of \textit{monotone properties} on $\basis{A}$. Specializing the
definition of monotonicity to algebraic CPO $A$ and codomain $\prop$,
we see that a property $P : \basis{A} \rightarrow \prop$ is monotone
when:
\[ x \: \sqsubseteq y \Rightarrow P \: x \Rightarrow P \: y. \]



I.e., $P$ is monotone when its holding on some approximation $x$
implies its holding on all further refinements of $x$ (all $y$ such
that $x \sqsubseteq y$). What then does it mean for a property to be
continuous? Recalling that a function $f : A \rightarrow B$ is
$\omega$-continuous when for every $\omega$-chain $C
: \nat \rightarrow A$, $f \: (\sup{C}) \simeq \sup{(f \circ C)}$,
specialized to $\co{P} : A \rightarrow \prop$:
\[ \co{P} \: (\sup{C}) \iff \sup{(\co{P} \circ C)} \]


\noindent
or, equivalently (by Definition~\ref{def:dense} and
Remark~\ref{remark:prop-lattice}):
\begin{align}
  \co{P} \: a \iff \exists i. \: P \: (\mathsf{idl} \: a \: i).
  \label{eq:continuous-property}
\end{align}

We see that the property $\co{P}$ holds on element $a : A$ if and only
if its restriction to $\basis{A}$ holds for \textit{some}
approximation of $a$. This leads us to the definition of continuous
property:

\begin{definition}[Continuous property]
  \label{def:continuous-property}
  
  Let $A$ be an algebraic CPO. A predicate $P : \pred{A}$ is
  a \textit{continuous property} on $A$ when for all $a : A$:
  \[ P \: a \iff \exists i. \: P \: (\mathsf{incl} \: (\mathsf{idl} \:
  a \; i)). \]
\end{definition}


It follows that continuous extensions of decidable properties are
semi-decidable. We also have the dual notion of \textit{cocontinuous
properties}:


\begin{definition}[Cocontinuous property]
  \label{def:cocontinuous-property}
  
  Let $A$ be an algebraic CPO. A predicate $P : \pred{A}$ is
  a \textit{cocontinuous property} on $A$ when for all $a : A$:
  \[ P \: a \iff \forall i. \: P \: (\mathsf{incl} \: (\mathsf{idl} \:
  a \; i)). \]
\end{definition}




\section{Conats}
\label{sec:conats}



Our first concrete example of a coinductive algebraic CPO is the type
$\conat$ of \textit{conats}, the natural numbers extended with a
``point at infinity'' $\omega_\nat$.

\begin{definition}[\link{conat}{\cwpconat}{$\conat$ ($\mathsf{conat}$)}]
  \label{def:conat}
  
  Define the type $\conat$ of conats coinductively by the formation
  rules:
  \begin{mathpar}
    \inferrule [\link{conat}{\cwpcozero}{conat-zero}]
               { }
               { \cons{cozero} : \conat }
    \and
    \inferrule [\link{conat}{\cwpcosucc}{conat-succ}]
               { n : \conat }
               { \cons{cosucc} \: n : \conat }
  \end{mathpar}
\end{definition}

\begin{definition}[\link{conat}{\cwpomega}{$\omega_{\nat}$}]
  \label{ex:omega-nat}
  
  Define the infinite conat $\omega_{\nat}$ coinductively:
  \[ \omega_{\nat} \triangleq \cons{cosucc}{} \: \omega_{\nat} \]
\end{definition}

The order relation on $\conat$ is the usual ordering of natural
numbers extended so that $n \sqsubseteq \omega_\nat$ for all $n
: \conat$.
To obviate the need for excessive reliance of generalized
rewriting~\cite{DBLP:journals/jfrea/Sozeau09}, we introduce an
extensionality axiom entailing propositional equality from order
equivalence. The same is done for coinductive types of streams, trees,
and tries in Sections~\ref{sec:colists}, \ref{sec:cotries},
and \ref{sec:cotrees}, respectively (see
Appendix~\ref{app:coinductive-extensionality} for discussion).

\begin{axiom}[\link{conat}{\cwpconatzext}{Conat extensionality}]
  \label{axiom:conat-extensionality}

  $\forall \: n \: m : \conat, \: n \simeq_{\conat} m \Rightarrow n = m$.
\end{axiom}



The type $\nat$ of natural numbers serves as a compact basis for
$\conat$, where the inclusion
map \link{conat}{\cwpnatzinj}{$\mathsf{incl_{\nat,\conat}}
: \nat \rightarrow \conat$} injects natural numbers into $\conat$,
and \link{conat}{\cwpconatzprefix}{$\mathsf{idl_{\nat,\conat}}
: \conat \rightarrow \nat \rightarrow \nat$} generates convergent
chains of finite approximations of conats.








\begin{remark}[\link{conat}{\cwpCompactznat}{$\nat$ is compact}]
  \label{remark:nat-compact}
\end{remark}

\begin{remark}[\link{conat}{\cwpaCPOzconat}{$\nat$ is dense in $\conat$}]
  \label{remark:nat-dense}

  For all $n : \conat$,
  \begin{align*}
    \mathsf{idl} \: n \text{ is an $\omega$-chain}, and \\
    \sup{(\mathsf{incl} \circ \mathsf{idl} \: n)} = n.
  \end{align*}
\end{remark}

\begin{remark}[\link{colist}{\cwpaCPOzcolist}{$\conat$ is a pointed algebraic CPO with bottom element $\cons{cozero}$ and basis $\nat$}]
  \label{remark:conat-aCPO}
\end{remark}

\subsection{Unlimited Fuel}
\label{subsec:infinite-fuel}




When a function is not inductive on the structure of any of its
arguments, a common trick is to define it instead by induction on a
separate $\nat$ argument (the \textit{fuel}), and ensure that enough
fuel is always provided for the function to complete its task. We
define fueled computations via the following `$\mathsf{iter}$'
construction that, starting from initial element $z$, repeatedly
applies a function $f$ until exhausting the fuel.



\begin{definition}[\link{conat}{\cwpnatziter}{$\mathsf{iter}$}]
  \label{def:iter}

  For type $A$, $z : A$, $f : A \rightarrow A$, and $n : \nat$, define
  $\mathsf{iter} \: z \: f \: n : A$ by induction on $n$:
  \begin{center}
    \setlength{\tabcolsep}{3pt}
    \begin{tabular}{l l l}
      \rowcolor{lightgray}
      \multicolumn{3}{l}{$\mathsf{iter} \: z \: f : \nat \rightarrow
      A$} \\
      \hline
      $\cons{O}$ & $\triangleq$ & $z$ \\
      $\cons{S} \: n$ & $\triangleq$ & $f \: (\mathsf{iter} \: f \:
      n)$
    \end{tabular}
  \end{center}
\end{definition}

By taking the continuous extension of a fueled iteration, we extend
its domain to include $\omega_\nat$, allowing it to be supplied an
unlimited amount of fuel!

\begin{definition}[\link{conat}{\cwpcoiter}{$\mathsf{coiter}$}]
  \label{def:coiter}

  For pointed type $A$ and $f : A \rightarrow A$, define
  $\mathsf{coiter} \: f : \conat \rightarrow A$ by:
  \[ \mathsf{coiter} \: f \triangleq \co{(\mathsf{iter} \: \bot_A \:
  f)}. \]
\end{definition}

We call this technique \textit{lazy coiteration}, and use it to
implement the Kleene closure operator on a coinductive encoding of
regular languages in Section~\ref{subsec:regular-languages}.

\section{Streams}
\label{sec:colists}


In this section we define \textit{streams} (or \textit{colists}) as a
coinductive algebraic CPO (Definition~\ref{def:colists}), define a
number of essential operations and predicates over them (including the
oft-problematic `filter' operation) (Section~\ref{subsec:cofolds}),
and illustrate the use of the basic proof principles of AlgCo to
reason about streams (Section~\ref{subsec:proving-with-fusion}),
culminating in the verification of a coinductive variant of the sieve
of Eratosthenes (Section~\ref{subsec:sieve-of-eratosthenes}).

\paragraph*{Streams as an Algebraic CPO}

Our definition of coinductive lists deviates slightly from the
standard definition of streams (e.g.,~\cite{chlipala2022certified}) by
inclusion of a bottom element:

\begin{definition}[\link{colist}{\cwpcolist}{Streams}]
  \label{def:colists}
  
  Define the type $\colist{A}$ of streams with element type $A$
  coinductively by the formation rules:
  \begin{mathpar}
    \inferrule [\link{colist}{\cwpconil}{stream-bot}]
               { }
               { \bot_{\colist{A}} : \colist{A} }
    \and
    \inferrule [\link{colist}{\cwpcocons}{stream-cons}]
               { a : A \\ l : \colist{A} }
               { \cons{cocons} \: a \: l : \colist{A} }
  \end{mathpar}
\end{definition}


We are careful not to regard $\bot_{\colist{A}}$ as simply a
$\cons{nil}$ constructor for streams, viewing it instead as
the \textit{undefined} or \textit{divergent} stream which loops
forever producing no output. Streams are ordered by a straightforward
structural prefix relation:

\begin{definition}[\link{colist}{\cwpcolistzle}{Stream order}]
  \label{def:colist-order}
  
  Define $\sqsubseteq_{\colist{A}} : \rel{\colist{A}}{\colist{A}}$
  coinductively by the inference rules:
  \begin{mathpar}
    \inferrule [\link{colist}{\cwpcolistzleznil}{$\sqsubseteq_{\colist{A}}$-bot}]
               { l : \colist{A} }
               { \bot_{\colist{A}} \sqsubseteq_{\colist{A}} l }
    \and
    \inferrule [\link{colist}{\cwpcolistzlezcons}{$\sqsubseteq_{\colist{A}}$-cons}]
               { a : A \\ l_1 \sqsubseteq_{\colist{A}} l_2 }
               { \cons{cocons} \: a \: l_1 \sqsubseteq_{\colist{A}} \cons{cocons} \: a \: l_2 }
  \end{mathpar}
\end{definition}

Intuitively, we have $l_1 \sqsubseteq_{\colist{A}} l_2$ when either
$l_1$ is a $\bot$-terminated finite approximation of $l_2$, or $l_1$
and $l_2$ are equal. A compact basis for $\colist{A}$ is given by the
inductive type $\alist{A}$ of lists (with constructors $\cons{nil}$
and $\cons{cons}$) with prefix ordering:

  

  

\begin{definition}[\link{colist}{\cwplistzle}{List order}]
  \label{def:list-order}
  
  Define $\sqsubseteq_{\alist{A}} : \rel{\alist{A}}{\alist{A}}$
  inductively by the inference rules:
  \begin{mathpar}
    \inferrule [\link{colist}{\cwplistzleznil}{$\sqsubseteq_{\alist{A}}$-nil}]
               { l : \alist{A} }
               { \cons{nil} \sqsubseteq_{\alist{A}} l }
    \and
    \inferrule [\link{colist}{\cwplistzlezcons}{$\sqsubseteq_{\alist{A}}$-cons}]
               { a : A \\ l_1 \sqsubseteq_{\alist{A}} l_2 }
               { \cons{cons} \: a \: l_1 \sqsubseteq_{\alist{A}} \cons{cons} \: a \: l_2 }
  \end{mathpar}
\end{definition}


\begin{remark}[\link{colist}{\cwpCompactzlist}{$\alist{A}$ is compact}]
  \label{remark:list-compact}
\end{remark}

\begin{remark}[\link{colist}{\cwpaCPOzcolist}{$\alist{A}$ is dense in $\colist{A}$}]
  \label{remark:colist-dense}

  Let $A$ be a type. Then, for all $a : \colist{A}$, $\mathsf{idl} \:
  a$ is an $\omega$-chain and
  $\sup{(\mathsf{incl} \circ \mathsf{idl} \: a)} = a$.

\end{remark}

Remarks~\ref{remark:list-compact} and~\ref{remark:colist-dense} show
that $\alist{A}$ is a compact basis for $\colist{A}$. Furthermore, we
have $\bot_{\colist{A}} \sqsubseteq l$ for all $l : \colist{A}$, and
thus:

\begin{remark}[\link{colist}{\cwpaCPOzcolist}{$\colist{A}$ is a pointed algebraic CPO with basis $\alist{A}$}]
  \label{remark:colist-aCPO}
\end{remark}




\begin{example}[\link{sieve}{\cwpnats}{$\mathsf{nats}$}]
  \label{ex:nats}
  
  For $n : \nat$, define the stream $\mathsf{nats} \: n
  : \colist{\nat}$ of natural numbers starting from $n$ coinductively
  by:
  \[ \mathsf{nats} \: n \triangleq \cons{cocons} \: n \:
  (\mathsf{nats} \: (\cons{S} \: n)) \]
\end{example}

\subsection{Cofolds}
\label{subsec:cofolds}

Many continuous extensions over streams share a common computational
structure. In this section we define an abstraction over this common
pattern (``cofolds'') and use it to derive definitions and computation
rules for standard operations on streams. In
Section~\ref{subsec:extracting-sieve} we provide an extraction
primitive for cofolds for lazy execution in Haskell. The basis of
cofolds is the standard right-associative fold operator on lists:

\begin{definition}[\link{colist}{\cwpfold}{$\mathsf{fold}$}]
  \label{def:fold}

  For types $A$ and $B$, $z : B$, $f : A \rightarrow B \rightarrow B$,
  and $l : \alist{A}$, define $\mathsf{fold} \: z \: f \: l : B$ by
  induction on $l$:
  \begin{center}
    \setlength{\tabcolsep}{3pt}
    \begin{tabular}{l l l}
      \rowcolor{lightgray}
      \multicolumn{3}{l}{$\mathsf{fold} \: z \: f
      : \alist{A} \rightarrow B$} \\
      \hline
      $\cons{nil}$ & $\triangleq$ & $z$ \\
      $\cons{cons} \: a \: l$ & $\triangleq$ & $f \: a \:
      (\mathsf{fold} \: f \: l)$
    \end{tabular}
  \end{center}
\end{definition}

Operations of the form $\mathsf{fold} \: z \: f$ are often
called \textit{catamorphisms}~\cite{meijer1991functional}, or
simply \textit{folds}. We introduce \textit{cofolds}: continuous
functions on streams of the form $\co{(\mathsf{fold} \: \bot \: f)}$
for monotone $f$, and \textit{anticofolds} (written
$\mathsf{\hat{co}fold}$): cocontinuous functions of the form
$\coop{(\mathsf{fold} \: \top \: f)}$ for antimonotone $f$.


\begin{definition}[\link{colist}{\cwpcofold}{$\mathsf{cofold}$}]
  \label{def:cofold}

  For type $A$, pointed type $B$, and $f : A \rightarrow B \rightarrow
  B$, define $\mathsf{cofold} \: f : \colist{A} \rightarrow B$ by:
  \[ \mathsf{cofold} \: f \triangleq \co{(\mathsf{fold} \: \bot_B \:
  f)} \]
  or when $B$ has a top element $\top_B$, define
  $\mathsf{\hat{co}fold} \: f : \colist{A} \rightarrow B$:
  \[ \mathsf{\hat{co}fold} \:
  f \triangleq \coop{(\mathsf{fold} \: \top_B \: f)}. \]
\end{definition}

\link{colist}{\cwpmonotonezfold}{$\mathsf{cofold} \: f$
is continuous when $f$ is monotone,}
and \link{colist}{\cwpantimonotonezfold}{$\mathsf{\hat{co}fold} \: f$
is cocontinuous when $f$ is antimonotone.}

\paragraph*{Cofold computation}
The following generic computation lemma can be used to derive
computation rules for cofolds.

\begin{lemma}[\link{colist}{\cwpcofoldzcons}{$\mathsf{cofold}$ computation}]
  \label{lemma:cofold-computation}

  Let $A$ be a type, $B$ an ordered type, and $f : A \rightarrow
  B \rightarrow B$. Then, if $B$ is a pointed CPO and $f \: a$ is
  continuous for every $a$,
  \[ \mathsf{cofold} \: f \: (\cons{cocons} \: a \: l) \simeq f \:
  a \: (\mathsf{cofold} \: f \: l) \]
  or if $B$ is an lCPO with a top element and $f \: a$ is
  l-continuous for every $a$,
  \[ \mathsf{\hat{co}fold} \: f \: (\cons{cocons} \: a \: l) \simeq
  f \: a \: (\mathsf{\hat{co}fold} \: f \: l). \]
\end{lemma}

Recall that `$\simeq$` stands for order-equivalence
(Definition~\ref{def:order-equivalence}), which often implies
propositional equality (e.g., for $\eR$ by antisymmetry and $\conat$
by Axiom~\ref{axiom:conat-extensionality}).

\paragraph*{Example cofolds}
Here we present some illustrative cofolds over streams and derive
their computation rules from Lemma~\ref{lemma:cofold-computation}. We
sometimes give explicit names to the basis functions being extended to
make proofs about them more readable (e.g.,
Definition~\ref{def:cofilter} and
Lemma~\ref{lemma:cofilter-comm}). Our first example is the cofold
$\mathsf{length_{\colist{A}}}$ mapping cotrees to $\conat$ such that
the length of any infinite stream is equal to
$\mathsf{\omega_{\nat}}$.





\begin{definition}[\link{colist}{\cwpcolistzlength}{$\mathsf{length_{\colist{A}}}$}]
  \label{def:colength}

  For type $A$, define $\mathsf{length_{\colist{A}}}
  : \colist{A} \rightarrow \co{\nat} \triangleq \co{(\mathsf{fold} \: \cons{cozero} \:
  (\lambda \_. \: \cons{cosucc}))}$,
  with \link{colist}{\cwpcosumzcons}{computation rule:}
  \[ \mathsf{length_{\colist{A}}} \: (\cons{cocons} \: \_ \: l)
  = \cons{cosucc} \: (\mathsf{length_{\colist{A}}} \: l). \]
\end{definition}




\paragraph*{Quantifying Predicates Over Streams}

Existential quantification over streams is given by the continuous
predicate:



\begin{definition}[\link{colist}{\cwpcolistzexists}{$\co{\exists_P}$}]
  \label{def:coexists}

  For type $A$ and predicate $P : A \rightarrow \prop$, define
  $\co{\exists_P}
  : \colist{A} \rightarrow \prop \triangleq \co{(\mathsf{fold} \: \bot \:
  (\lambda a. \: \lambda Q. \: P \: a \lor Q))}$,
  with derived introduction and elimination rules:
  \begin{mathpar}
  \inferrule [\link{colist}{\cwpcolistzexistszintrow}{$\co{\exists_P}$-intro-1}]
             { P \: a \\ l : \colist{A} }
             { \co{\exists_P} \: (\cons{cocons} \: a \: l) }
  \and
  \inferrule [\link{colist}{\cwpcolistzexistszintroe}{$\co{\exists_P}$-intro-2}]
             { a : A \\ \co{\exists_P} \: l }
             { \co{\exists_P} \: (\cons{cocons} \: a \: l) }
  \and
  \inferrule [\link{colist}{\cwpcolistzexistszelim}{$\co{\exists_P}$-elim}]
             { \co{\exists_P} \: (\cons{cocons} \: a \: l) }
             { P \: a \lor \co{\exists_P} \: l }
  \end{mathpar}
\end{definition}



\noindent and universal quantification by the cocontinuous
predicate:



\begin{definition}[\link{colist}{\cwpcolistzforall}{$\coop{\forall_P}$}]
  \label{def:coforall}

  For type $A$ and predicate $P : A \rightarrow \prop$, define
  $\coop{\forall_P}
  : \colist{A} \rightarrow \prop \triangleq \coop{(\mathsf{fold} \: \top \:
  (\lambda a. \: \lambda Q. \: P \: a \land Q))}$,
  with derived introduction and elimination rules:
  \begin{mathpar}
  \inferrule [\link{colist}{\cwpcolistzforallzintro}{$\coop{\forall_P}$-intro}]
             { P \: a \\ \coop{\forall_P} \: l }
             { \coop{\forall_P} \: (\cons{cocons} \: a \: l) }
  \and
  \inferrule [\link{colist}{\cwpcolistzforallzelimw}{$\coop{\forall_P}$-elim-1}]
             { \coop{\forall_P} \: (\cons{cocons} \: a \: l) }
             { P \: a }
  \and
  \inferrule [\link{colist}{\cwpcolistzforallzelime}{$\coop{\forall_P}$-elim-2}]
             { \coop{\forall_P} \: (\cons{cocons} \: a \: l) }
             { \coop{\forall_P} \: l }
  \end{mathpar}
\end{definition}

Definitions~\ref{def:coexists} and~\ref{def:coforall} are used to
prove the correctness properties of the sieve of Eratosthenes in
Theorems~\ref{theorem:sieve-complete} and~\ref{theorem:sieve-sound}.

\paragraph*{Order Relation as Continuous Extension}
Although a primitive order relation on streams must have already been
defined (Definition~\ref{def:colist-order}) to gain access to the
machinery of the AlgCo framework in the first place, we can re-define
it as a continuous extension and prove it equivalent to the original.

\begin{definition}[\link{colist}{\cwpcolistzlex}{$\coop{\sqsubseteq_{\alist{A}}}$}]
  \label{def:co-le}

  For type $A$, define $\coop{\sqsubseteq_{\alist{A}}}
  : \colist{A} \rightarrow \colist{A} \rightarrow \prop$ where
  $\link{colist}{\cwplistzcolistzle}{\sqsubseteq_{\alist{A}}:}
  \alist{A} \rightarrow \colist{A} \triangleq$
  \begin{align*}
    & \mathsf{fold} \: (\lambda \_. \: \top) \: (\lambda a. \: \lambda
    f. \: \lambda l. \text{ match } l \text{ with } \\
    & \hspace{104pt} | \: \bot_{\colist{A}} \Rightarrow \bot \\
    & \hspace{104pt} | \: \cons{cocons} \: b \: l' \Rightarrow a =
    b \land f \: l' \\
    & \hspace{105pt} \text{end})
  \end{align*}
\end{definition}

\begin{remark}[\link{colist}{\cwpcolistzlezcolistzlex}{$\coop{\sqsubseteq_{\alist{A}}}$ coincides with $\sqsubseteq_{\colist{A}}$}]
  \label{remark:cole-coincides}

  Let $A$ be a type. Then, $\forall \: l_1 \: l_2 : \colist{A}, \:
  l_1 \coop{\sqsubseteq_{\alist{A}}} l_2 \iff
  l_1 \sqsubseteq_{\colist{A}} l_2.$
\end{remark}

Definition~\ref{def:co-le} is often more convenient in practice than
the primitive order relation (Definition~\ref{def:colist-order}) due
to being fuseable (by Theorem~\ref{theorem:fusion}) with other
continuous extensions.

\paragraph*{Cofilter}
Filtering a coinductive stream by a given predicate is a notoriously
awkward exercise, typically requiring the stream type to be extended
with a special constructor for so-called ``silent-steps'' (as in,
e.g.,~\cite{xia2019interaction}, inducing significant definitional
clutter and performance overhead), or the use of complicated mixed
well-founded induction-coinduction schemes as
in~\cite{bertot2005filters}. We easily define filter as a cofold:

\begin{definition}[\link{colist}{\cwpcofilter}{$\mathsf{filter_{\colist{A}}}$}]
  \label{def:cofilter}

  For type $A$ and $f : A \rightarrow \bool$, define
  $\mathsf{filter_{\colist{A}}} \: f
  : \colist{A} \rightarrow \colist{A} \triangleq \co{(\mathsf{filter_{\alist{A}}} \:
  f)}$, where:
  \[ \link{colist}{\cwpafilter}{\mathsf{filter_{\alist{A}}}} \:
  f \triangleq \mathsf{fold} \: \bot \: (\lambda a. \: \lambda
  l. \: \text{if } f \: a \text{ then } \cons{cocons} \: a \:
  l \text{ else } l) \]
  with \link{colist}{\cwpcofilterzcons}{computation rule:}
  \begin{align*}
    & \mathsf{filter_{\colist{A}}} \: f \: (\cons{cocons} \: a \: l) = \\
    & \hspace{15pt} \text{if} f \: a \text{ then } \cons{cocons} \:
    a \: (\mathsf{filter_{\colist{A}}} \: f \: l) \text{ else
    } \mathsf{filter_{\colist{A}}} \: f \: l.
  \end{align*}
\end{definition}


  


\subsection{Proving with Fusion}
\label{subsec:proving-with-fusion}


  

Here we illustrate the use of AlgCo to prove commutativity of
$\mathsf{filter_{\colist{A}}}$. First we use
Theorem~\ref{theorem:fusion} to fuse both sides of the equation, and
then we apply Lemma~\ref{lemma:equivalence-continuous-extension} to
reduce the proof to induction over basis elements.

\begin{lemma}[\link{colist}{\cwpcofilterzcomm}{$\mathsf{filter_{\colist{A}}}$ is commutative}]
  \label{lemma:cofilter-comm}
  
  Let $A$ be a type and $f$, $g : A \rightarrow \bool$. Then,
  \[ \forall \: l : \colist{A}, \mathsf{filter_{\colist{A}}} \: f \:
  (\mathsf{filter_{\colist{A}}} \: g \: l)
  = \mathsf{filter_{\colist{A}}} \: g \:
  (\mathsf{filter_{\colist{A}}} \: f \: l). \]
  Proof.
  Unfolding the definition of $\mathsf{filter_{\colist{A}}}$, we wish
  to show:
  \[ \mathsf{filter_{\colist{A}}} \:
  f \: \circ \co{(\mathsf{filter_{\alist{A}}} \: g)}
  = \mathsf{filter_{\colist{A}}} \:
  g \: \circ \co{(\mathsf{filter_{\alist{A}}} \: f)}. \]
  We first fuse both sides of the equation by
  Theorem~\ref{theorem:fusion}:
  \[ \co{(\mathsf{filter_{\colist{A}}} \:
  f \circ \mathsf{filter_{\alist{A}}} \: g)}
  = \co{(\mathsf{filter_{\colist{A}}} \:
  g \circ \mathsf{filter_{\alist{A}}} \: f)} \]
  which then follows by
  Corollary~\ref{lemma:equivalence-continuous-extension} from:
  \[ \forall \: l : \alist{A}, \mathsf{filter_{\colist{A}}} \: f \:
  (\mathsf{filter_{\alist{A}}} \: g \: l)
  = \mathsf{filter_{\colist{A}}} \: g \:
  (\mathsf{filter_{\alist{A}}} \: f \: l) \]
  which follows by straightforward induction on $l$ and application of
  the $\mathsf{filter_{\colist{A}}}$ computation rule
  (Definition~\ref{def:cofilter}). $\qed$
\end{lemma}

\subsection{Proving (Co-)Continuous Properties}
\label{subsec:proving-cocontinuous-properties}

A surprising but elegant feature of the AlgCo framework is that by
considering predicates over algebraic CPOs as a special case of
continuous functions, we can easily compose them with continuous
extensions and greatly simplify their proofs via fusion. To
illustrate, let us prove that $\mathsf{filter_{\colist{A}}} \: P$
preserves $\coop{\forall_Q}$ for any $P$ and $Q$:

\begin{lemma}[\link{colist}{\cwpcolistzforallzcofilterx}{$\mathsf{filter_{\colist{A}}}$ preserves $\coop{\forall_Q}$}]
  \label{lemma:cofilter-forall}

  Let $A$ be a type, $P : \pred{A}$, $Q : \pred{A}$, and $s
  : \colist{A}$ such that $\coop{\forall_Q} \: s$. Then,
  $\coop{\forall_Q} \: (\mathsf{filter_{\colist{A}}} \: P \: s)$.

  Proof. Unfolding the definition of $\mathsf{filter_{\colist{A}}}$,
  the goal becomes:
  \[ \coop{\forall_Q} \: (\co{(\mathsf{filter_{\alist{A}}} \: P)} \:
  s). \]
  By \link{aCPO}{\cwpcozcoopP}{fusion} this is equivalent to:
  \[ \coop{(\forall_Q \circ \mathsf{filter_{\alist{A}}} \: P)} \:
  s \]
  which
  by \link{aCPO}{\cwpcoopzintro}{\textsf{$\mathsf{\hat{co}}$-intro}}
  (Definition~\ref{def:cocontinuous-property}) follows from:
  \[ \forall i : \nat, \: \forall_Q \: (\mathsf{filter_{\alist{A}}} \:
  P \: (\mathsf{idl} \: s \: i)). \]
  Fix $i$. By the fact that $\forall l : \alist{A}, \forall_Q \:
  l \Rightarrow \coop{\forall_Q} \: (\mathsf{filter_{\alist{A}}} \:
  P \: l)$ (\link{colist}{\cwplistzforallzforallzafilter}{by
  straightforward induction on $l$}), it suffices to show:
  \[ \forall_Q \: (\mathsf{idl} \: s \: i) \]
  which follows
  by \link{aCPO}{\cwpcoopzintro}{\textsf{$\mathsf{\hat{co}}$-elim}}
  (Definition~\ref{def:cocontinuous-property}) from
  $\coop{\forall_Q} \: s. \qed$
\end{lemma}

\subsection{Sieve of Eratosthenes}
\label{subsec:sieve-of-eratosthenes}


The sieve of Eratosthenes is an ancient algorithm for generating a
sequence of prime numbers up to a given limit. An infinitary variant
of the sieve (sometimes called ``the unfaithful
sieve''~\cite{o2009genuine}), while often used to demonstrate the
elegance of lazy functional programming, is difficult to replicate in
a sound type theory (such as Coq's) due to its use of the filter
operation on streams. We define the sieve as a monotone function on
lists continuously extended to streams:


\begin{definition}[\link{sieve}{\cwpsievezaux}{$\mathsf{sieve}$}]
  \label{def:sieve}

  Define $\mathsf{sieve}
  : \colist{\mathbb{Z}} \triangleq \mathsf{sieve\_aux} \:
  (\mathsf{nats} \: 2)$, where:
  \[ \link{sieve}{\cwpsievezaux}{\mathsf{sieve\_aux}}
  : \colist{\mathbb{Z}} \rightarrow \colist{\mathbb{Z}} \triangleq \co{\mathsf{sieve\_aux_{\alist{A}}}} \textit{, where} \]
  \begin{align*}
  & \link{sieve}{\cwpasievezaux}{\mathsf{sieve\_aux_{\alist{A}}}} \triangleq \\
  & \mathsf{fold} \: \bot_{\colist{\mathbb{Z}}} \: (\lambda
  n. \: \lambda l. \: \cons{cocons} \: n \: (\mathsf{filter_{\colist{A}}} \:
  (\lambda m. \: m \: \mathsf{mod} \: n \neq 0) \: l))
  \end{align*}
  with \link{sieve}{\cwpsievezcons}{computation rule:}
  \begin{align*}
  & \mathsf{sieve\_aux} \: (\cons{cocons} \: n \: l) = \\
  & \hspace{15pt} \cons{cocons} \: n \: (\mathsf{filter_{\colist{A}}} \:
  (\lambda m. \: m \: \mathsf{mod} \: n \neq
  0) \: (\mathsf{sieve\_aux} \: l)).
  \end{align*}
\end{definition}




We prove that $\mathsf{sieve}$ is complete
(Theorem~\ref{theorem:sieve-complete}) and sound
(Theorem~\ref{theorem:sieve-sound}) wrt. the prime numbers, and that
it generates them in ascending order without duplicates
(Theorem~\ref{theorem:sieve-sorted-nodup}).


\begin{definition}[$\mathsf{is\_prime}$]
  \label{def:is-prime}
  Define $\mathsf{is\_prime} : \mathbb{Z} \rightarrow \prop$ by:
  \[ \mathsf{is\_prime} \: n \triangleq 1 < n \land \forall \: m, \: 1
  < m \rightarrow n \neq m \rightarrow n \: \mathsf{mod} \: m \neq
  0. \]
\end{definition}

\begin{theorem}[\link{sieve}{\cwpsievezcomplete}{$\mathsf{sieve}$ is complete wrt. the prime numbers}]
  \label{theorem:sieve-complete}

  Let $n : \mathbb{Z}$ such that $\mathsf{is\_prime} \: n$. Then,
  $\co{\exists_{\mathsf{eq} \: n}} \: \mathsf{sieve}$. I.e., every
  prime number $n$ appears in the stream generated by
  $\mathsf{sieve}$.
\end{theorem}


Theorem~\ref{theorem:sieve-complete} states a continuous property
(Definition~\ref{def:continuous-property}) of the sieve stream, and so
is proved by showing for every $n$ that it holds for \textit{some}
finite approximation of the sieve.

\begin{theorem}[\link{sieve}{\cwpsievezsound}{$\mathsf{sieve}$ is sound wrt. the prime numbers}]
  \label{theorem:sieve-sound}

  $\coop{\forall_{\mathsf{is\_prime}}} \: \mathsf{sieve}$. I.e., every
  number appearing in the stream generated by $\mathsf{sieve}$ is
  prime.
\end{theorem}


Theorem~\ref{theorem:sieve-sound} states a cocontinuous property
(Definition~\ref{def:cocontinuous-property}) of the sieve stream, and
so is proved by showing that it holds for \textit{all} finite
approximations of the sieve.

\begin{theorem}[\link{sieve}{\cwpsortedzsieve}{$\mathsf{sieve}$ is sorted} and \link{sieve}{\cwpnodupzsieve}{contains no duplicates}]
  \label{theorem:sieve-sorted-nodup}
\end{theorem}

Theorem~\ref{theorem:sieve-sorted-nodup} is proved by showing that the
stream $\mathsf{nats} \: 2$ is strictly increasing and that the
continuous extension $\mathsf{sieve\_aux}$
(Definition~\ref{def:sieve}) is strictly order preserving.

\subsection{Extracting the Sieve}
\label{subsec:extracting-sieve}




Continuous extensions are noncomputable because $\mathsf{sup}$
(Section~\ref{subsec:domain-theory}) is nonconstructive.
Nevertheless, it is possible to implement an extraction primitive for
cofolds over streams (shown in Figure~\ref{fig:cofold-haskell}) for
lazy execution in Haskell. The correctness of the cofold extraction
primitive is justified by the cofold computation rule in
Lemma~\ref{lemma:cofold-computation}, the specialization of which can
be checked explicitly on streams one intends to execute (e.g., the
sieve computation rule in Definition~\ref{def:sieve}).



\paragraph*{On Computability of Continuous Extensions}


The extraction primitive shown in Figure~\ref{fig:cofold-haskell}
provides a computational interpretation of cofolds that is only
partially correct because programs extracted from continuous
extensions are not guaranteed to terminate for every input (cf. the
extracted fixpoint operators of~\cite{DBLP:conf/itp/Chargueraud10}
and~\cite{DBLP:conf/ppdp/BertotK08}). The reason is that continuous
predicates are only \textit{semi-decidable} (see
Section~\ref{subsec:cocontinuous-properties}), and continuity in
general corresponds only to \textit{partial computability}. For
example, the continuous Boolean-valued predicate
\link{colist}{\cwpbadzbool}{$\mathsf{bad}$} $\triangleq \co{(\lambda \_ \: x. \: x)}
: \colist{\bool} \rightarrow \bool$ diverges for every stream.



We may, however, define a notion of \textit{productivity} of streams
generated by continuous extensions (the proof of which for our sieve
follows from Euclid's theorem on
the \link{inf\_primes}{\cwpexzprimezgt}{infinitude of
primes}~\cite{heath1956thirteen}):

\lstset{
 language=haskell,
 columns=[c]fixed,
 basicstyle=\small\ttfamily,
 keywordstyle=\bfseries,
 commentstyle=\color{OliveGreen},  
 breaklines=true,
 showstringspaces=false}

\begin{figure}
  \centering
  \begin{tabular}{c}
\begin{lstlisting}
cofold =
  \ o p f l ->
    case l of
      Conil -> bot o p
      Cocons a l' -> f a (cofold o p f l')
\end{lstlisting}
  \end{tabular}

  \caption{Haskell extraction primitive for
  $\mathsf{cofold}$. Parameters $o$ and $p$
  are \link{order}{\cwpOType}{$\mathsf{OType}$}
  and \link{order}{\cwpPType}{$\mathsf{PType}$} instance dictionary
  objects for the order relation of the codomain.}

  \label{fig:cofold-haskell}
\end{figure}

\begin{definition}[\link{colist}{\cwpproductivexx}{productive}]
  \label{def:productive}

  For any type $A$, a stream $s : \colist{A}$ is said to be
  (infinitely) \textit{productive} when
  $\mathsf{length_{\colist{A}}} \: s = \omega_{\nat}$.
\end{definition}


We prove that $\mathsf{sieve}$ is productive:

\begin{theorem}[\link{sieve}{\cwpproductivezsieve}{$\mathsf{sieve}$ is productive}]
  \label{theorem:sieve-productive}

  $\mathsf{length_{\colist{\mathbb{Z}}}} \: \mathsf{sieve}
  = \omega_{\nat}$.
\end{theorem}

But this notion of productivity is not sufficient in general to
absolutely guarantee the absence of divergence. To see why, consider
the stream defined coinductively by
\link{colist}{\cwpbadzstream}{$\mathsf{bad\_stream}$} $\triangleq \cons{cocons} \: (\mathsf{bad} \:
(\mathsf{nats} \: 0)) \: \mathsf{bad\_stream}$. Each element of
$\mathsf{bad\_stream}$ is \link{colist}{\cwpbadzstreamzspec}{provably
equal to $\false$},
and \link{colist}{\cwpbadzstreamzproductive}{$\mathsf{bad\_stream}$ is
productive,} but any attempt to extract and compute with its elements
will immediately diverge.



\section{Coinductive Tries}
\label{sec:cotries}



Formal languages, typically defined as sets of strings over an
alphabet $\Sigma$, can be given an elegant representation by infinite
prefix trees (\textit{cotries}) branching over $\Sigma$. The cotrie
representation of a formal language encodes an automaton for
recognizing strings in the language by marking each node as either
accepting or rejecting the empty string $\epsilon$ and directly
encoding the Brzozowski derivative at the node wrt.~each symbol of the
alphabet via a coinductive
continuation~\cite{DBLP:phd/dnb/Traytel15}. In this section, we define
the type of cotries as an algebraic CPO, develop regular operations
over them, and prove that the operations satisfy the axioms of a
Kleene algebra, resulting in extraction of verified regular expression
matchers to Haskell (see, e.g., the file
$\href{https://github.com/bagnalla/algco/blob/main/RE_test.v}{RE\_test.v}$).

\paragraph*{Formal languages as an algebraic CPO}

The type of cotries is defined with respect to a finite alphabet type
(indexing the children of each node) and a pointed ordered type for
node labels. For ease of presentation, we specialize to a fixed finite
alphabet $\Sigma$ and label type $\bool$.

  

\begin{figure}[t]
  \centering
  \begin{tikzpicture}
	\begin{pgfonlayer}{nodelayer}
		\node [style={green_circle}] (0) at (-7.75, 4.75) {};
		\node [style=none] (7) at (-15.25, 0.75) {};
		\node [style=none] (15) at (-15.25, 0.25) {\tiny{...}};
		\node [style=none] (23) at (-9.75, 4.5) {\tiny{a}};
		\node [style=none] (24) at (-5.75, 4.5) {\tiny{b}};
		\node [style=none] (25) at (-12.75, 3.5) {\tiny{a}};
		\node [style=none] (26) at (-5, 3.5) {\tiny{a}};
		\node [style=none] (27) at (-10.5, 3.5) {\tiny{b}};
		\node [style=none] (28) at (-2.75, 3.5) {\tiny{b}};
		\node [style=none] (48) at (-14.25, 0.75) {};
		\node [style=none] (49) at (-14.25, 0.25) {\tiny{...}};
		\node [style=none] (52) at (-13.25, 0.75) {};
		\node [style=none] (53) at (-13.25, 0.25) {\tiny{...}};
		\node [style=none] (56) at (-12.25, 0.75) {};
		\node [style=none] (57) at (-12.25, 0.25) {\tiny{...}};
		\node [style=none] (60) at (-11.25, 0.75) {};
		\node [style=none] (61) at (-11.25, 0.25) {\tiny{...}};
		\node [style=none] (64) at (-10.25, 0.75) {};
		\node [style=none] (65) at (-10.25, 0.25) {\tiny{...}};
		\node [style=none] (68) at (-9.25, 0.75) {};
		\node [style=none] (69) at (-9.25, 0.25) {\tiny{...}};
		\node [style=none] (72) at (-8.25, 0.75) {};
		\node [style=none] (73) at (-8.25, 0.25) {\tiny{...}};
		\node [style={red_circle}] (123) at (-14.75, 1.75) {};
		\node [style={green_circle}] (124) at (-12.75, 1.75) {};
		\node [style={red_circle}] (125) at (-10.75, 1.75) {};
		\node [style={red_circle}] (126) at (-8.75, 1.75) {};
		\node [style={red_circle}] (131) at (-13.75, 2.75) {};
		\node [style={green_circle}] (132) at (-9.75, 2.75) {};
		\node [style={green_circle}] (135) at (-11.75, 3.75) {};
		\node [style=none] (136) at (-7.25, 0.75) {};
		\node [style=none] (137) at (-7.25, 0.25) {\tiny{...}};
		\node [style=none] (141) at (-6.25, 0.75) {};
		\node [style=none] (142) at (-6.25, 0.25) {\tiny{...}};
		\node [style=none] (145) at (-5.25, 0.75) {};
		\node [style=none] (146) at (-5.25, 0.25) {\tiny{...}};
		\node [style=none] (149) at (-4.25, 0.75) {};
		\node [style=none] (150) at (-4.25, 0.25) {\tiny{...}};
		\node [style=none] (153) at (-3.25, 0.75) {};
		\node [style=none] (154) at (-3.25, 0.25) {\tiny{...}};
		\node [style=none] (157) at (-2.25, 0.75) {};
		\node [style=none] (158) at (-2.25, 0.25) {\tiny{...}};
		\node [style=none] (161) at (-1.25, 0.75) {};
		\node [style=none] (162) at (-1.25, 0.25) {\tiny{...}};
		\node [style=none] (165) at (-0.25, 0.75) {};
		\node [style=none] (166) at (-0.25, 0.25) {\tiny{...}};
		\node [style={red_circle}] (176) at (-6.75, 1.75) {};
		\node [style={red_circle}] (177) at (-4.75, 1.75) {};
		\node [style={green_circle}] (178) at (-2.75, 1.75) {};
		\node [style={red_circle}] (179) at (-0.75, 1.75) {};
		\node [style={green_circle}] (180) at (-5.75, 2.75) {};
		\node [style={red_circle}] (181) at (-1.75, 2.75) {};
		\node [style={green_circle}] (182) at (-3.75, 3.75) {};
		\node [style=none] (183) at (-14.5, 2.5) {\tiny{a}};
		\node [style=none] (184) at (-15.25, 1.25) {\tiny{a}};
		\node [style=none] (185) at (-13.25, 1.25) {\tiny{a}};
		\node [style=none] (186) at (-10.5, 2.5) {\tiny{a}};
		\node [style=none] (187) at (-11.25, 1.25) {\tiny{a}};
		\node [style=none] (188) at (-9.25, 1.25) {\tiny{a}};
		\node [style=none] (189) at (-6.5, 2.5) {\tiny{a}};
		\node [style=none] (190) at (-7.25, 1.25) {\tiny{a}};
		\node [style=none] (191) at (-5.25, 1.25) {\tiny{a}};
		\node [style=none] (192) at (-2.5, 2.5) {\tiny{a}};
		\node [style=none] (193) at (-3.25, 1.25) {\tiny{a}};
		\node [style=none] (194) at (-1.25, 1.25) {\tiny{a}};
		\node [style=none] (195) at (-9, 2.5) {\tiny{b}};
		\node [style=none] (196) at (-13, 2.5) {\tiny{b}};
		\node [style=none] (197) at (-12.25, 1.25) {\tiny{b}};
		\node [style=none] (198) at (-14.25, 1.25) {\tiny{b}};
		\node [style=none] (199) at (-10.25, 1.25) {\tiny{b}};
		\node [style=none] (200) at (-8.25, 1.25) {\tiny{b}};
		\node [style=none] (201) at (-6.25, 1.25) {\tiny{b}};
		\node [style=none] (202) at (-5, 2.5) {\tiny{b}};
		\node [style=none] (203) at (-4.25, 1.25) {\tiny{b}};
		\node [style=none] (204) at (-1, 2.5) {\tiny{b}};
		\node [style=none] (205) at (-2.25, 1.25) {\tiny{b}};
		\node [style=none] (206) at (-0.25, 1.25) {\tiny{b}};
		\node [style=none] (207) at (-7.75, 5.75) {};
	\end{pgfonlayer}
	\begin{pgfonlayer}{edgelayer}
		\draw [style=arrow] (0) to (135);
		\draw [style=arrow] (135) to (131);
		\draw [style=arrow] (131) to (123);
		\draw [style=arrow] (131) to (124);
		\draw [style=arrow] (135) to (132);
		\draw [style=arrow] (132) to (125);
		\draw [style=arrow] (132) to (126);
		\draw [style=arrow] (182) to (180);
		\draw [style=arrow] (180) to (176);
		\draw [style=arrow] (180) to (177);
		\draw [style=arrow] (182) to (181);
		\draw [style=arrow] (181) to (178);
		\draw [style=arrow] (181) to (179);
		\draw [style=arrow] (0) to (182);
		\draw [style=arrow] (123) to (7.center);
		\draw [style=arrow] (123) to (48.center);
		\draw [style=arrow] (124) to (52.center);
		\draw [style=arrow] (124) to (56.center);
		\draw [style=arrow] (125) to (60.center);
		\draw [style=arrow] (125) to (64.center);
		\draw [style=arrow] (126) to (68.center);
		\draw [style=arrow] (126) to (72.center);
		\draw [style=arrow] (176) to (136.center);
		\draw [style=arrow] (176) to (141.center);
		\draw [style=arrow] (177) to (145.center);
		\draw [style=arrow] (177) to (149.center);
		\draw [style=arrow] (178) to (153.center);
		\draw [style=arrow] (178) to (157.center);
		\draw [style=arrow] (179) to (161.center);
		\draw [style=arrow] (179) to (165.center);
		\draw [style=arrow] (207.center) to (0);
	\end{pgfonlayer}
\end{tikzpicture}
  \caption{Coinductive trie encoding of regular language `$\epsilon + a^*b + b^*a$' over alphabet \{a, b\}. Green nodes indicate accept states of the encoded automaton.}
  \label{fig:trie}
\end{figure}

\begin{definition}[\link{cotrie}{\cwpcotrie}{$\lang$}]
  \label{def:lang}
  
  Define the type $\lang$ of languages over finite alphabet $\Sigma$
  coinductively by the formation rule:
  \begin{mathpar}
    \inferrule [\link{cotrie}{\cwpcotrieznode}{lang-node}]
               { b : \bool \\ \forall \: a : \Sigma, \: k \: a : \lang }
               { \cons{lnode} \: b \: k : \lang }
  \end{mathpar}
\end{definition}

  

\begin{definition}[\link{cotrie}{\cwpcotriezle}{$\lang$ order}]
  \label{def:lang-order}
  
  Define $\sqsubseteq_{\lang} : \rel{\lang}{\lang}$ coinductively by
  the inference rule:
  \begin{mathpar}
    \inferrule [\link{cotrie}{\cwpcotriezleznode}{$\sqsubseteq_{\lang}$-node}]
               { b_1 \sqsubseteq_{\bool} b_2 \\
                 \forall \: a : \Sigma, \: f \: a \sqsubseteq_{\lang} g \: a }
               { \cons{lnode} \: b_1 \: f \sqsubseteq_{\lang} \cons{lnode} \: b_2 \: g }
  \end{mathpar}
\end{definition}


Unlike the other coinductive types in this paper, there is no
constructor for $\bot$, or any finite constructor at all (provided
that the alphabet is nonempty). Every cotrie is thus infinite and
congruent in structure to every other cotrie, differing only by the
values of node labels. The empty language in which every label is
false serves as the bottom element:

  

\begin{definition}[\link{cotrie}{\cwpcotriezbot}{$\emptyset$ language}]
  \label{def:bot-cotrie}
  
  Define the empty language $\emptyset$ coinductively by:
  $\emptyset \triangleq \cons{lnode} \: \false \:
  (\lambda \_. \: \emptyset)$.
\end{definition}



Although all cotries are infinite, we can take as a compact basis the
subset of cotries that are meaningfully defined only up to a finite
depth (being equal to $\emptyset$ thereafter). Such cotries are
finitely approximable and thus compact, and can be represented by an
inductive type with a $\bot$ constructor standing in for $\emptyset$:

  


\begin{definition}[\link{cotrie}{\cwptrie}{$\tlang$}]
  \label{def:tlang}
  
  Define the type $\tlang$ of finite trie languages over alphabet
  $\Sigma$ inductively by the formation rules:
  \begin{mathpar}
    \inferrule [\link{cotrie}{\cwptriezbot}{tlang-bot}]
               { }
               { \bot_{\tlang} : \tlang }
    \and
    \inferrule [\link{cotrie}{\cwptrieznode}{trie-node}]
               { b : \bool \\ \forall \: a : \Sigma, \: k \: a : \tlang }
               { \cons{tlnode} \: k : \tlang }
  \end{mathpar}
\end{definition}



However, if we were to define the order relation on $\tlang$ in the
usual way (e.g., Definition~\ref{def:colist-order}), the inclusion map
into $\lang$ would not be injective (up to order equivalence), as
multiple elements of $\tlang$ (all approximations of $\emptyset$)
would be mapped to $\emptyset$. We solve this by defining the order
relation in such a way that all approximations of the bottom cotrie
are collapsed into a single equivalence class.

  

\begin{definition}[\link{cotrie}{\cwptriezle}{$\tlang$ order}]
  \label{def:tlang-order}
  
  Define $\sqsubseteq_{\tlang} : \rel{\tlang}{\tlang}$ inductively by
  the inference rules:
  \begin{mathpar}
    \inferrule [\link{cotrie}{\cwptriezlezbot}{$\sqsubseteq_{\tlang}$-bot}]
               { \mathsf{is\_bot} \: t_1 \\ t_2 : \tlang }
               { t_1 \sqsubseteq_{\tlang} t_2 }
    \and
    \inferrule [\link{cotrie}{\cwptriezleznode}{$\sqsubseteq_{\tlang}$-node}]
               { b_1 \sqsubseteq b_2 \\
                 \forall \: a : \Sigma, \: f \: a \sqsubseteq_{\tlang} g \: a }
               { \cons{tlnode} \: b_1 \: f \sqsubseteq_{\tlang} \cons{tlnode} \: b_2 \: g }
  \end{mathpar}
  where \link{cotrie}{\cwpiszbot}{$\mathsf{is\_bot}
  : \tlang \rightarrow \prop$} is defined inductively by:
  \begin{mathpar}
    \inferrule [\link{cotrie}{\cwpiszbotzbot}{$\mathsf{is\_bot}$-bot}]
               { }
               { \mathsf{is\_bot} \: \bot_{\tlang} }
    \and
    \inferrule [\link{cotrie}{\cwpiszbotznode}{$\mathsf{is\_bot}$-node}]
               { \forall \: a : \Sigma, \mathsf{is\_bot} \: (k \: a) }
               { \mathsf{is\_bot} \: (\cons{tlnode} \: \false \: k) }
  \end{mathpar}
\end{definition}




\begin{remark}[\link{cotrie}{\cwpCompactztrie}{$\tlang$ is compact}]
  \label{remark:tlang-compact}
\end{remark}

\begin{remark}[\link{cotrie}{\cwpaCPOzcotrie}{$\lang$ is a pointed algebraic CPO with basis $\tlang$}]
  \label{remark:lang-aCPO}
\end{remark}




\subsection{Regular Languages}
\label{subsec:regular-languages}




\begin{definition}[$\lang$ membership]
  \label{def:in-lang}
  
  For $t : \lang$ and $l : \alist{\Sigma}$, define
  $\mathsf{in\_lang} \: t \: l : \bool$ by induction on $l$:
  \begin{center}
    \setlength{\tabcolsep}{3pt}
    \begin{tabular}{l l l l}
      \rowcolor{lightgray}
      \multicolumn{4}{l}{$\mathsf{in\_lang}
      : \lang \rightarrow \alist{\Sigma} \rightarrow \bool$} \\
      \hline
      ($\cons{lnode} \: b \: \_$) & $\cons{nil}$ & $\triangleq$ &
      $b$ \\
      ($\cons{lnode} \: \_ \: k$) & $(\cons{cons} \: a \: l)$ &
      $\triangleq$ & $\mathsf{in\_lang} \: (k \: a) \: l$
    \end{tabular}
  \end{center}
\end{definition}


A remarkable property of $\lang$ is that the semantic notion of
language equivalence coincides exactly with structural equality
(cf. the extensional tries
of~\cite{DBLP:journals/corr/abs-2110-05063}). This greatly simplifies
proofs of equations between cotries by reducing them to
straightforward induction on lists.

\begin{lemma}[\link{cotrie}{\cwpinzlangzcotriezle}{Semantic equivalence coincides with structural equality}]
  \label{lemma:semantic-structural}
  
  Let $a : \lang$ and $b : \lang$. $\Sigma$. Then, $(\forall l
  : \alist{\Sigma}, \: \mathsf{in\_lang} \: a \: l
  = \mathsf{in\_lang} \: b \: l) \iff a = b$.
\end{lemma}


We let `$\mathsf{o} \: a$' denote the Boolean indicating whether $a$
accepts the empty string or not and we let `$\delta_a \: x$' denote
the Brzozowski derivative of language $a$ wrt.~character $x$ (i.e.,
$\mathsf{o} \: (\cons{lnode}{} \: b \: \_) \triangleq b$ for all $b$
and $\delta_{(\cons{lnode \: \_ \: k)}{}} \: x \triangleq k \: x$ for
all $k$ and $x$).

  

\begin{definition}[Regular language constructions]
  \label{def:regular-language-constructions} Define the language
  $\epsilon$ containing the empty string:
  \begin{align*}
    & \link{cotrie}{\cwpepsilon}{\epsilon} \triangleq \cons{lnode} \: \true \:
    (\lambda \_. \: \emptyset)
  \end{align*}
  and define the union, intersection, and complement of languages
  via primitive coinduction:
  \begin{align*}
    & \link{cotrie}{\cwpunion}{a + b} \triangleq \cons{lnode} \:
    (\mathsf{o} \: a \lor \mathsf{o} \: b) \: (\lambda
    x. \: \delta_a \: x + \delta_b \: x) \\
    & \link{cotrie}{\cwpintersection}{a \: \& \:
    b} \triangleq \cons{lnode} \: (\mathsf{o} \: a \land \mathsf{o} \:
    b) \: (\lambda x. \: \delta_a \: x \: \& \: \delta_b \: x) \\
    & \link{cotrie}{\cwpcomplement}{\lnot
    a} \triangleq \cons{lnode} \: (\lnot (\mathsf{o} \: a)) \:
    (\lambda x. \: \lnot (\delta_a \: x))
  \end{align*}
\end{definition}

The structural order on cotrie languages also coincides with the
standard semantic order on Kleene
algebras: \link{cotrie}{\cwpcotriezlezorder}{$a \sqsubseteq b \iff a +
b = a$.} The concatenation and Kleene star operators are more
difficult to define than those given above, because they are not
primitive corecursive. We define concatenation as the continuous
extension of a monotone fold over finite tries:

\begin{definition}[\link{cotrie}{\cwpconcat}{$\mathsf{concat}$}]
  \label{def:concat}

  For languages $a : \lang$ and $b : \lang$, define the concatenation
  $a \cdot b : \lang \triangleq \co{\mathsf{tconcat}} \: a \: b$,
  where:
  \begin{align*}
  & \link{cotrie}{\cwptconcat}{\mathsf{tconcat}} \triangleq \mathsf{fold_{\tlang}} \:
  (\lambda \_. \: \emptyset) \\
  & \hspace{20pt} (\lambda b \: k \: l. \: \cons{lnode} \: (b \land
  o \: l) \\
  & \hspace{40pt} (\lambda x. \: k \: x \: l + (\text{if } b \text{
  then } \delta_l \: x \text{ else } \emptyset)))
  \end{align*}
\end{definition}

The Kleene star is a lazy coiteration (see
Section~\ref{subsec:infinite-fuel}):

\begin{definition}[\link{cotrie}{\cwpstar}{Kleene star}]
  \label{def:kleene-star}

  For language $a : \lang$, define the Kleene closure $a^* : \lang$
  by:
  \[ a^* \triangleq \mathsf{coiter} \: (\lambda
  b. \: \cons{lnode} \: \true \: (\delta_a \: x \cdot
  b)) \: \omega_{\nat}\]
\end{definition}

All the operations defined in this section can be extracted to Haskell
and used to decide membership of strings over a finite alphabet in
regular languages. We prove finally that they satisfy the Kleene
algebra axioms:

\begin{theorem}[\link{cotrie}{\cwpKleeneAlgebraLawszlang}{Kleene Algebra}]
  \label{theorem:kleene-algebra}

  The type $\lang$ with zero element $\emptyset$, one element
  $\epsilon$, and $+$, $\cdot$, and $^*$ operations forms a Kleene
  algebra.
\end{theorem}




\section{Coinductive Binary Trees}
\label{sec:cotrees}


In this section, we define the coinductive type of infinitary trees
with finite branching factor (\textit{cotrees}) as an algebraic CPO.
We then use cotrees to encode samplers for discrete distributions in
the random bit model~\cite{von195113,saad2020sampling} and show how
AlgCo enables expected value
~\cite{DBLP:journals/jcss/Kozen85,DBLP:journals/toplas/MorganMS96,DBLP:phd/dnb/Kaminski19}
reasoning about them
(Section~\ref{subsec:discrete-distribution-samplers}), culminating in
Theorem~\ref{theorem:equidistribution} showing that sequences of
samples are \textit{equidistributed}~\cite{BecherG22} wrt.~the
expected value semantics of the cotrees used to generate them. The
results of this section are used to verify samplers compiled from
probabilistic programs in the \zar
system~\cite{https://doi.org/10.48550/arxiv.2211.06747}.

\paragraph*{Coinductive Trees as an Algebraic CPO}
For clarity of presentation we specialize cotrees to the case of
coinductive \textit{binary} trees with index type $\bool$ (the index
type must be finite in general to ensure the compactness of basis
elements).

\begin{definition}[\link{cotree}{\cwpcotree}{Cotrees}]
  \label{def:cotrees}
  
  Define the type $\cotree{A}$ of cotrees with element type $A$
  coinductively by the formation rules:
  \begin{mathpar}
    \inferrule [\link{cotree}{\cwpcobot}{cotree-bot}]
               { }
               { \bot_{\cotree{A}} : \cotree{A} }
    \and
    \inferrule [\link{cotree}{\cwpcoleaf}{cotree-leaf}]
               { a : A  }
               { \cons{coleaf}{} \: a : \cotree{A} }
    \and
    \inferrule [\link{cotree}{\cwpconode}{cotree-node}]
               { \forall \: b : \bool, \: k \: b : \cotree{A} }
               { \cons{conode}{} \: k : \cotree{A} }
  \end{mathpar}
\end{definition}

\noindent and straightforward structural prefix relation:

\begin{definition}[\link{cotree}{\cwpcotreezle}{Cotree order}]
  \label{def:cotree-order}
  
  Define $\sqsubseteq_{\cotree{A}} : \rel{\cotree{A}}{\cotree{A}}$
  coinductively by the inference rules:
  \begin{mathpar}
    \inferrule [\link{cotree}{\cwpcotreezlezbot}{$\sqsubseteq_{\cotree{A}}$-bot}]
               { l : \cotree{A} }
               { \bot_{\cotree{A}} \sqsubseteq_{\cotree{A}} l }
    \and
    \inferrule [\link{cotree}{\cwpcotreezlezleaf}{$\sqsubseteq_{\cotree{A}}$-leaf}]
               { a : A }
               { \cons{coleaf} \: a \sqsubseteq_{\cotree{A}} \cons{coleaf} \: a }
    \and
    \inferrule [\link{cotree}{\cwpcotreezleznode}{$\sqsubseteq_{\cotree{A}}$-node}]
               { \forall \: b : \bool, \: f \: b \sqsubseteq_{\cotree{A}} g \: b }
               { \cons{conode} \: f \sqsubseteq_{\cotree{A}} \cons{conode} \: g }
  \end{mathpar}
\end{definition}

A compact basis for cotrees is given by a corresponding ordered type
of finite binary trees (with order corresponding to that of cotrees
(Definition~\ref{def:cotree-order}), omitted for brevity):

\begin{definition}[\link{cotree}{\cwpatree}{Finite trees}]
  \label{def:alists}
  
  Define the type $\atree{A}$ of finite binary trees with element type
  $A$ inductively by the formation rules:
  \begin{mathpar}
    \inferrule [\link{cotree}{\cwpabot}{tree-bot}]
               { }
               { \bot_{\atree{A}} : \atree{A} }
    \and
    \inferrule [\link{cotree}{\cwpaleaf}{tree-leaf}]
               { a : A  }
               { \cons{leaf} \: a : \atree{A} }
    \and
    \inferrule [\link{cotree}{\cwpanode}{tree-node}]
               { \forall \: b : \bool, \: k \: b : \atree{A} }
               { \cons{node} \: k : \atree{A} }
  \end{mathpar}
\end{definition}

  

\begin{remark}[\link{cotree}{\cwpCompactzatree}{$\atree{A}$ is compact}]
  \label{remark:atree-compact}
\end{remark}

\begin{remark}[\link{cotree}{\cwpaCPOzcotree}{$\cotree{A}$ is a pointed algebraic CPO with basis $\atree{A}$}]
  \label{remark:cotree-aCPO}
\end{remark}

\link{cotree}{\cwptcofold}{Cofolds over cotrees} and their \link{cotree}{\cwptcofoldzleaf}{computation rules} are derived analogously to streams (Section~\ref{subsec:cofolds}) wrt.~the fold operator:

\begin{definition}[\link{cotree}{\cwptfold}{$\mathsf{fold_{\atree{A}}}$}]
  \label{def:tfold}

  For type $A$, ordered type $B$, $z : B$, $f : A \rightarrow B$, $g :
  (\bool \rightarrow B) \rightarrow B$, and $t : \atree{A}$, define
  $\mathsf{fold_{\atree{A}}} \: z \: f \: g \: t : B$ by induction on
  $t$:
  \begin{center}
    \setlength{\tabcolsep}{3pt}
    \begin{tabular}{l l l}
      \rowcolor{lightgray}
      \multicolumn{3}{l}{$\mathsf{fold_{\atree{A}}} \: z \: f \: g
      : \atree{A} \rightarrow B$} \\
      \hline
      $\bot_{\atree{A}}$ & $\triangleq$ & $z$ \\
      $\cons{leaf} \: a$ & $\triangleq$ & $f \: a$ \\
      $\cons{node} \: k$ & $\triangleq$ & $h \:
      (\mathsf{fold_{\atree{A}}} \: z \: f \: g \circ k)$
    \end{tabular}
  \end{center}
\end{definition}

\subsection{Discrete Distribution Samplers}
\label{subsec:discrete-distribution-samplers}

A cotree $t : \cotree{A}$ can be interpreted as a sampling procedure
over sample space $A$ in the random bit model, where it is provided a
stream of uniformly distributed random bits from which to generate
samples. Indeed, that is the approach taken in the
Zar~\cite{https://doi.org/10.48550/arxiv.2211.06747} system, in which
probabilistic programs with unbounded loops and conditioning are
compiled to cotree representations of samplers over their posterior
distributions.

The sampler encoded by a cotree $t : \cotree{A}$ can be understood as
follows: starting from the root, consume random bits from the
environment to guide a traversal of $t$ by consuming one bit at each
$\cons{conode}$ to determine whether to take the left or right
subtree. The process terminates when (if ever) a $\cons{coleaf} \: x$
is reached, yielding sample $x : A$.

For example, consider the cotree sampler $t_{\frac{2}{3}}$ in
Figure~\ref{fig:two-thirds} encoding a Bernoulli distribution with
$p=\frac{2}{3}$ (the probability of $\true$). The process encoded by
$t_{\frac{2}{3}}$ terminates with probability $1$ (i.e., it is
``almost surely'' terminating). However, although we can be confident
that it will terminate eventually, we cannot place any finite upper
bound on how many steps it will take because there technically exists
an infinite execution path of alternating right and left
choices. 

\begin{figure}[t]
  \centering
  \begin{tikzpicture}
	\begin{pgfonlayer}{nodelayer}
		\node [style=circle] (0) at (-4.25, 4.5) {};
		\node [style=square] (1) at (-6.25, 3.5) {$\true$};
		\node [style=circle] (2) at (-2.25, 3.5) {};
		\node [style={red_square}] (3) at (-0.25, 2.5) {$\false$};
		\node [style=none] (4) at (-4.25, 5.5) {};
		\node [style=circle] (5) at (-4.25, 2.5) {};
		\node [style=circle] (6) at (-2.25, 1.5) {};
		\node [style=square] (7) at (-6.25, 1.5) {$\true$};
		\node [style={red_square}] (9) at (-0.25, 0.5) {$\false$};
		\node [style=none] (14) at (-4.25, 0.5) {};
		\node [style=none] (15) at (-4.5, 0.25) {...};
	\end{pgfonlayer}
	\begin{pgfonlayer}{edgelayer}
		\draw [style=arrow] (0) to (1);
		\draw [style=arrow] (4.center) to (0);
		\draw [style=arrow] (2) to (3);
		\draw [style=arrow] (0) to (2);
		\draw [style=arrow] (2) to (5);
		\draw [style=arrow] (5) to (6);
		\draw [style=arrow] (5) to (7);
		\draw [style=arrow] (6) to (9);
		\draw [style=arrow] (6) to (14.center);
	\end{pgfonlayer}
\end{tikzpicture}
  \caption{Cotree encoding of Bernoulli($\frac{2}{3}$) distribution
  over sample space $\bool$.}
  \label{fig:two-thirds}
\end{figure}

To reason formally about cotree samplers, we define a variant of
the \textit{weakest pre-expectation} ($\mathsf{wp}$) semantics
(originally due to Kozen~\cite{DBLP:journals/jcss/Kozen85}), adapted
from its application to the \textit{probabilistic Guarded Command
Language} ($\mathsf{pGCL}$)~\cite{DBLP:series/mcs/McIverM05}.

An \textit{expectation} is a function $f : A \rightarrow \eR$ mapping
elements of sample space $A$ to the nonnegative extended reals. The
purpose of $\mathsf{wp}$ is to compute expected values of expectations
over cotrees:
\begin{definition}[\link{cocwp}{\cwpcotwp}{$\mathsf{wp_{\cotree{}}}$}]
  \label{def:cowp}

  For type $A$ and expectation $f : A \rightarrow \eR$, define
  $\mathsf{wp_{\cotree{A}}} \: f
  : \cotree{A} \rightarrow \eR \triangleq \co{(\mathsf{wp_{\atree{A}}} \:
  f)}$, where:
  \[ \link{cocwp}{\cwpbtwp}{\mathsf{wp_{\atree{A}}}} \:
  f \triangleq \mathsf{fold_{\atree{A}}} \: 0 \: f \: (\lambda
  k. \: \frac{k \: \true + k \: \false}{2}) \]
  with \link{cocwp\_facts}{\cwpcotwpzleaf}{computation rules:}
  \begin{align*}
    & \mathsf{wp_{\cotree{A}}} \: f \: \bot_{\cotree{A}} = 0 \\
    & \mathsf{wp_{\cotree{A}}} \: f \: (\cons{coleaf} \: a) = f \:
    a \\
    & \mathsf{wp_{\cotree{A}}} \: f \: (\cons{conode} \: k)
    = \frac{\mathsf{wp_{\cotree{A}}} \: f \: (k \: \true)
    + \mathsf{wp_{\cotree{A}}} \: f \: (k \: \false)}{2}.
  \end{align*}
\end{definition}

$\mathsf{wp_{\cotree{A}}}$ can be used to express the probability of a
given event $Q : A \rightarrow \prop$ over the distribution encoded by
cotree $t : \cotree{A}$ via the expected value of the indicator
function $[Q]$ given by $\mathsf{wp_{\cotree{A}}} \: [Q] \: t$. For
example, the probability that $t_\frac{2}{3}$ produces the value
$\true$ is given by $\mathsf{wp} \: [\lambda x. \: x = \true] \:
t_{\frac{2}{3}} = \frac{2}{3}$. Technically,
$\mathsf{wp_{\cotree{A}}} \: [Q] \: t$ denotes the probability
of \textit{terminating with a sample satisfying $Q$}, and does not
account for executions which produce no sample at all.






The following lemma shows a fundamental connection between
$\mathsf{wp_{\cotree{}}}$ and \link{cotree}{\cwpcotreezbind}{monadic
bind} (in the $\cotree{}$ monad) that is especially relevant when
reasoning about cotree samplers compiled from probabilistic programs
with sequenced commands:

\begin{lemma}[\link{cocwp\_facts}{\cwpcotwpzbind}{$\mathsf{wp_{\cotree{}}}$ bind}]
  \label{lemma:cowp-bind}

  Let $A$ and $B$ be types, $f : B \rightarrow \eR$, $t : \cotree{A}$,
  and $k : A \rightarrow \cotree{B}$. Then,
  \[ \mathsf{wp_{\cotree{B}}} \: f \: (t \bind k)
  = \mathsf{wp_{\cotree{A}}} \: (\mathsf{wp_{\cotree{B}}} f \circ
  k) \: t. \]
\end{lemma}


We further validate $\mathsf{wp_{\cotree{}}}$ by proving several
healthiness conditions, including Markov's
inequality~\cite{DBLP:phd/dnb/Kaminski19}. All proofs are carried out
with the machinery of AlgCo from
Section~\ref{sec:algebraic-coinductives}.

\begin{theorem}[\link{cocwp\_facts}{\cwpmarkovzinequality}{Markov's inequality}]
  \label{lemma:markov-inequality}

  Let $A$ be a type, $f : A \rightarrow \eR$, $t : \cotree{A}$, and $a
  : \Rpos$. Then,
  \[ \mathsf{wp_{\cotree{A}}} \: [f \ge a] \:
  t \le \frac{\mathsf{wp_{\cotree{A}}} \: f \: t}{a} \]
\end{theorem}

\subsection{Coinductive Measure}
\label{subsec:coinductive-measure}

A cotree sampler can alternatively be viewed as an encoding of a
partial function mapping infinite bitstreams (elements of
the \textit{Cantor space} $2^\omega$) to samples. Given $t
: \cotree{A}$, we let $f_t : 2^\omega \rightharpoonup A$ denote the
function induced by $t$ that either diverges on a given bitstream or
produces a sample $x$ by starting from the root of $t$ and using the
bits of the stream to guide traversal (e.g., taking the left subtree
on $1$ and the right on $0$) until reaching a leaf containing $x$.

The \textit{preimage} $f_t^{-1}(Q)$ of an event $Q$ under sampler
function $f_t$ is the subset of bitstreams in $2^\omega$ sent by $f_t$
to samples in $Q$. We represent subsets of $2^\omega$ as cotrees of
type $\cotree{\alist{\bool}}$ encoding countable unions
of \textit{basic sets} in $2^\omega$. Basic sets are encoded by finite
bitstrings, where bitstring $b$ denotes the set $\{ s : 2^\omega \mid
b \sqsubseteq s\}$ of all bitstreams sharing prefix $b$.  We further
require that all bitstrings appearing in a cotree set be pairwise
incomparable, i.e., disjoint. We let $\Sigma_1^0$ denote the class of
countable unions of basic sets, and show how to compute cotree
representations of $\Sigma_1^0$ preimages of events via a continuous
extension in Definition~\ref{def:preimage}.

Under this view, we re-cast the task of inferring the probability of
event $Q$ wrt.~sampler $t$ to that of computing the \textit{measure}
of $f_t^{-1}(Q) \subseteq 2^\omega$, where the measure of a bitstring
$bs$ is equal to $\frac{1}{2^{(length \: bs)}}$, and the measure of a
$\Sigma_1^0$ cotree (see Definition~\ref{def:tcosum}) is the sum of
the measures of its constituent bitstrings (so long as they are
pairwise disjoint). We prove that the probability of any event $Q$
according to the $\mathsf{wp_{\cotree{A}}}$ semantics of sampler $t$
coincides with the measure of its preimage under $f_t$
(Lemma~\ref{lemma:wp-measure-preimage}), and then use this to prove
that sequences of samples generated from $t$ are equidistributed
wrt.~its $\mathsf{wp_{\cotree{A}}}$ semantics
(Theorem~\ref{theorem:equidistribution}).

\paragraph*{Computing preimages}
Preimages of cotrees are defined by the composition of two continuous
extensions $\mathsf{lang_{\cotree{A}}}$ and
$\mathsf{filter_{\cotree{A}}}$. We begin with
$\mathsf{lang_{\cotree{A}}}$, which indiscriminately (not wrt.~any
predicate) computes the subset of $2^\omega$ sent by sampler $t$
to \textit{any leaf at all} (i.e., the preimage $f_t^{-1}(A)$ of the
entire sample space $A$, or, the \textit{language} of $t$).

\begin{definition}[\link{mu}{\cwpcotreezlang}{$\mathsf{lang_{\cotree{A}}}$}]
  \label{def:cotree-lang}

  For type $A$, define $\mathsf{lang_{\cotree{A}}}
  : \cotree{A} \rightarrow \cotree{\alist{\bool}} \triangleq \co{\mathsf{lang_{\atree{A}}}}$,
  where:
  \begin{align*}
  & \link{mu}{\cwpatreezlang}{\mathsf{lang_{\atree{A}}}} \triangleq \mathsf{fold_{\atree{A}}} \: \bot_{\cotree{\alist{\bool}}} \:
  (\lambda \_. \: \cons{coleaf}{} \: \cons{nil}{}) \\
  & \hspace{78pt} (\lambda g. \cons{conode}{} \: (\lambda
  b. \: \mathsf{map_{\cotree{A}}} \: (\cons{cons}{} \: b) \: (g \:
  b)))
  \end{align*}
  with \link{mu}{\cwpcotreezlangzleaf}{computation rules:}
  \begin{align*}
    & \mathsf{lang_{\cotree{A}}} \: f \: (\cons{coleaf} \: a)
    = \cons{coleaf} \: \cons{nil} \\
    & \mathsf{lang_{\cotree{A}}} \: f \: (\cons{conode} \: k) = \\
    & \hspace{15pt} \cons{conode} \: (\lambda
    b. \: \mathsf{map_{\cotree{A}}} \: (\cons{cons} \: b) \:
    (\mathsf{lang_{\cotree{A}}} \: (k \: b))).
  \end{align*}
\end{definition}

We then define a continuous extension for filtering:

\begin{definition}[\link{cotree}{\cwpcotreezfilter}{$\mathsf{filter_{\cotree{A}}}$}]
  \label{def:tcofilter}

  For type $A$ and predicate $P : A \rightarrow \bool$, define
  $\mathsf{filter_{\cotree{A}}} \: P
  : \cotree{A} \rightarrow \cotree{A} \triangleq \co{(\mathsf{filter_{\atree{A}}} \:
  P)}$, where:
  \begin{align*}
  & \link{cotree}{\cwpatreezcotreezfilter}{\mathsf{filter_{\atree{A}}}} \:
  P \triangleq \\
  & \hspace{15pt} \mathsf{fold_{\atree{A}}} \: \bot_{\cotree{A}} \: (\lambda
  a. \: \text{if } P \: a \: \text{ then } \cons{coleaf} \: a \text{
  else } \bot_{\cotree{A}}) \: \cons{conode}
  \end{align*}
  with \link{cotree}{\cwpcotreezfilterzleaf}{computation rules:}
  \begin{align*}
    & \mathsf{filter_{\cotree{A}}} \: P \: (\cons{coleaf} \: a)
    = \text{if } P \: a \text{ then } \: \cons{coleaf} \: a \text{
    else } \bot_{\cotree{A}} \\
    & \mathsf{filter_{\cotree{A}}} \: P \: (\cons{conode} \: k)
    = \cons{conode} \: (\mathsf{filter_{\cotree{A}}} \: P \circ k).
  \end{align*}
\end{definition}

Notice that, in contrast to $\mathsf{filter_{\colist{A}}}$
(Definition~\ref{def:cofilter}), elements removed by
$\mathsf{filter_{\cotree{A}}}$ are simply replaced by
$\bot_{\cotree{A}}$ rather than restructuring the tree to eliminate
them entirely. This is no problem because we never compute with
preimage sets -- they are only part of the correctness specification
for the ultimate equidistribution theorem
(Theorem~\ref{theorem:equidistribution}).

The preimage of a predicate $Q$ wrt.~sampler $t$ is then obtained as
the language of $t$ after being filtered by $Q$.

\begin{definition}[\link{mu}{\cwpcotreezpreimage}{$\mathsf{preimage_{\cotree{A}}}$}]
  \label{def:preimage}

  For type $A$ and predicate $Q : A \rightarrow \bool$, define
  $\mathsf{preimage_{\colist{A}}}
  : \cotree{A} \rightarrow \cotree{\alist{\bool}}$ as:
  \[ \mathsf{preimage_{\colist{A}}} \triangleq \mathsf{lang_{\cotree{A}}} \circ \mathsf{filter_{\cotree{A}}} \:
  Q \]
\end{definition}

We define the desired measure on $\Sigma_1^0$ sets as an application
of the $\mathsf{sum_{\cotree{A}}}$ operator for summing a function
over a cotree.

\begin{definition}[\link{equidistribution}{\cwptcosum}{$\mathsf{sum_{\cotree{A}}}$}]
  \label{def:tcosum}

  For type $A$ and $f : A \rightarrow \eR$, define
  $\mathsf{sum_{\cotree{A}}} \: f
  : \cotree{A} \rightarrow \eR \triangleq \co{\mathsf{sum_{\atree{A}}} \:
  f}$, where:
  \[ \link{mu}{\cwpasum}{\mathsf{sum_{\atree{A}}}} \triangleq \mathsf{fold_{\atree{A}}} \:
  0 \: f \: (\lambda g. \: g \: \true + g \: \false) \]  
  with \link{mu}{\cwptcosumzleaf}{computation rules:}
  \begin{align*}
    & \mathsf{sum_{\cotree{A}}} \: f \: (\cons{coleaf} \: a) = f \:
    x \\
    & \mathsf{sum_{\cotree{A}}} \: f \: (\cons{conode} \: k) = \\
    & \hspace{30pt} \mathsf{sum_{\cotree{A}}} \: f \: (k \: \true)
    + \mathsf{sum_{\cotree{A}}} \: f \: (k \: \false)
  \end{align*}
\end{definition}

\begin{definition}[\link{equidistribution}{\cwpmu}{$\mathsf{mu_{\cotree{\alist{\bool}}}}$}]
  \label{def:mu}

  Define $\mathsf{mu_{\cotree{\alist{\bool}}}}
  : \cotree{\alist{\bool}} \rightarrow \eR \triangleq \mathsf{sum_{\cotree{\alist{bool}}}} \:
  (\lambda l. \: \frac{1}{2^{(\mathsf{length_{\alist{\bool}}} \: l)}})$
\end{definition}

\paragraph*{Proving Equidistribution With AlgCo}

In this section we use AlgCo to prove that that any sequence of
samples produced by a cotree sampler $t$ is \textit{equidistributed}
wrt.~the $\mathsf{wp_{\cotree{A}}}$ semantics of $t$. At a high level,
our strategy is to ``push forward'' through the sampler the assumption
of uniform distribution of the source of random bits.  We first need
the following two lemmas:

\begin{lemma}[\link{cocwp\_facts}{\cwpcotwpztcosumzpreimage}{$\mathsf{wp_{\cotree{A}}} \: \mathsf{filter_{\cotree{A}}}$}]
  \label{lemma:wp-filter}
  Let $A$ be a type and $Q : A \rightarrow \bool$. Then,
  $\mathsf{wp_{\cotree{A}}} \: [Q]
  = \mathsf{wp_{\cotree{A}}} \: \mathbf{1} \circ \mathsf{filter_{\cotree{A}}} \:
  Q$.
\end{lemma}

\begin{lemma}[\link{mu}{\cwpcotwpzmuzlang}{$\mathsf{wp_{\cotree{A}}} \: \mathbf{1}$ equals $\mathsf{mu_{\cotree{A}}} \circ \mathsf{lang_{\cotree{A}}}$}]
  \label{lemma:wp-mu-lang}
  Let $A$ be a type and $Q : A \rightarrow \bool$. Then,
  $\mathsf{wp_{\cotree{A}}} \: \mathbf{1}
  = \mathsf{mu_{\cotree{A}}} \circ \mathsf{filter_{\cotree{A}}} \: Q$.
\end{lemma}

The proofs of Lemmas~\ref{lemma:wp-filter} and~\ref{lemma:wp-mu-lang}
both proceed by fusing the RHS to a single comorphism and applying
Lemma~\ref{lemma:equivalence-continuous-extension} to reduce the goal
to straightforward induction over finite trees. It immediately follows
that the probability of any event $Q$ according to the
$\mathsf{wp_{\cotree{A}}}$ semantics of sampler $t$ coincides with the
measure of the preimage of $Q$ under $f_t$:

\begin{lemma}[\link{mu}{\cwpcotwpztcosumzpreimage}{$\mathsf{wp_{\cotree{A}}}$ is measure of preimage}]
  \label{lemma:wp-measure-preimage}

  Let $A$ be a type, $Q : A \rightarrow \bool$, and $t
  : \cotree{A}$. Then,
  $\mathsf{wp_{\cotree{A}}} \: [Q] \: t
  = \mathsf{mu_{\cotree{\alist{\bool}}}} \:
  (\mathsf{preimage_{\cotree{A}}} \: Q \: t)$.
\end{lemma}

To specify uniform distribution of the source of random bits, we use a
variation of the classic notion of ``uniform distribution modulo
1''~\cite{kuipers2012uniform,BecherG22} generalized to $\Sigma_1^0$
subsets of $2^\nat$.

\begin{definition}[\link{equidistribution}{\cwpuniform}{$\Sigma_1^0$-u.d.}]
  \label{def:sigma01-ud}
  
  A sequence $\{x_i\}$ of bitstreams is \textit{$\Sigma_1^0$-uniformly
  distributed} (\textit{$\Sigma_1^0$-u.d.}) when for every $U
  : \Sigma_1^0$,
  $\lim_{n \to \infty}{\frac{1}{n} \sum_{i=0}^n{[\co{\exists_{=
  x_i}} \: U]}} = \mathsf{mu}_{\cotree{\alist{\bool}}} \: U$.
\end{definition}

In other words, a sequence of bitstreams is uniformly distributed when
every $\Sigma_1^0$ set gets its ``proper share'' of samples as the
number of samples goes to infinity.

\begin{theorem}[\link{equidistribution}{\cwpcotreezsampleszequidistributed}{Equidistribution of samples}]
  \label{theorem:equidistribution}
  
  Let $A$ be a type, $t : \cotree{A}$, $Q : A \rightarrow \bool$,
  $\{x_n\}$ a $\Sigma_1^0$-u.d. sequence of bitstreams, and
  $\{f_t(x_n)\}$ a sequence of samples obtained by mapping $f_t$ over
  $\{x_n\}$. Then, $\{f_t(x_n)\}$ is
  $\mathsf{wp_{\cotree{A}}}$-equidistributed wrt.~$t$:
  \[ \lim_{n \to \infty}{\frac{1}{n} \sum_{i=0}^n{[Q \: (f_t (x_i))]}}
  = \mathsf{wp_{\cotree{A}}} \: t \: [Q] \: \sigma \]

  Proof.
  Rewrite the RHS by Lemma~\ref{lemma:wp-measure-preimage} so
  the goal becomes:  
  \[ \lim_{n \to \infty}{\frac{1}{n} \sum_{i=0}^n{[Q \: (f_t (x_i))]}}
  = \mathsf{mu_{\cotree{\alist{\bool}}}} \:
  (\mathsf{preimage_{\cotree{A}}} \: Q \: t). \]
  
  Then let $U = \mathsf{preimage_{\cotree{A}}} \: Q \: t$ in the
  assumption of $\Sigma_1^0$-u.d. to obtain:
  \[ \lim_{n \to \infty}{\frac{1}{n} \sum_{i=0}^n{[\co{\exists_{=
  x_i}} \: \mathsf{preimage_{\cotree{A}}} \: Q \: t]}}
  = \mathsf{mu}_{\cotree{\alist{\bool}}} \: \mathsf{preimage_{\cotree{A}}} \:
  Q \: t \] 
  from which the goal immediately follows
  since $\link{equidistribution}{\cwpproduceszinzsigmaqw}{Q \: (f_t
  (x_i)) \iff \co{\exists_{=
  x_i}} \: \mathsf{preimage_{\cotree{A}}} \: Q \: t.} \qed$  
\end{theorem}

\section{Related Work}
\label{sec:related-work}

\paragraph*{Parameterized Coinduction (paco)}
The paco library~\cite{DBLP:conf/popl/HurNDV13} provides the most well
known generalization of primitive coinduction in Coq. Paco replaces
the syntactic guardedness condition for coinductive proofs with a
semantic one, leading to a greater degree of compositionality in
comparison to Coq's built in $\mathsf{cofix}$ mechanism, and
alleviation of some common pitfalls, e.g., misapplication of
coinduction hypotheses that are not checked until
QED-time. Coinductive properties defined using paco are still,
however, defined as \textit{greatest fixed points} of monotone
functionals, and thus suffer from the problems described in the
introduction when attempting to define functional mappings over
coinductive structures. With AlgCo, we take an entirely different
approach by carving out a subset of coinductive types (those forming
algebraic CPOs) for which coinductive reasoning can be replaced by
inductive reasoning. The $\mathsf{cawu}$ framework of
Pous~\cite{DBLP:conf/lics/Pous16} generalizes the theory of
parameterized coinduction and simplifies its treatment of
``up-to''~\cite{milner1989communication,sangiorgi1998bisimulation}
enhancements of bisimulation, but fundamentally stands in the same
relation as $\mathsf{paco}$ to the present paper.

\paragraph*{Copatterns}
Abel et al.~\cite{DBLP:conf/popl/AbelPTS13} present a type-theoretic
foundation for programming with infinite data such as streams via
observations (i.e., \textit{copattern} matching). As the categorical
dual to inductive pattern matching, which provides a general
\textit{elimination} scheme for inductive data, copattern matching provides a
general \textit{introduction} scheme for coinductive data. AlgCo fills
a gap in the middle by providing a \textit{continuous elimination
scheme for algebraic coinductives}
(Lemma~\ref{lemma:continuous-extension}). For example, the function
$\mathsf{sum} : \mathsf{Stream} \: \eR \rightarrow \eR$ from our
introduction can't be defined via copatterns because the codomain is
not coinductive, but it \textit{can} be defined as a continuous
extension because the domain is algebraic.


\paragraph*{Regular Coinductives}

Jeannin et al.~\cite{DBLP:journals/fuin/JeanninKS17a} extend OCaml
with support for ``regular'' (finitely representable) coinductives,
and equip them with a general elimination scheme parameterized by
user-specified solvers, allowing the programmer to define mappings
from regular coinductive structures into CPOs (e.g., mapping regular
infinitary $\lambda$ terms to sets of their free variables via a
fixpoint solver). Many coinductive structures we are interested in,
however, such as the cotree samplers generated from probabilistic
programs described in
Section~\ref{subsec:discrete-distribution-samplers}), are not always
finitely representable. The notion of algebraic CPO generalizes to a
broader class of coinductive structures while still providing a
general (albeit less computable) principle for elimination.

\paragraph*{Friendly Functions and Sized Types}
Syntactic guardedness conditions on primitive corecursive definitions
are relaxed in the Isabelle/HOL proof assistant via so-called
``friendly functions''~\cite{DBLP:conf/esop/BlanchetteBL0T17}, which
preserve productivity of their arguments. Corecursive calls are thus
allowed to appear under applications of friendly functions.
Sized types~\cite{DBLP:conf/popl/HughesPS96} have been implemented in
the Agda proof assistant~\cite{agda2022doc} to fulfill a similar role
in broadening the scope of allowable corecursive definitions, and have
been used to implement the concatenation and closure operators on
coinductive tries~\cite{DBLP:phd/dnb/Traytel15} as in
Section~\ref{subsec:regular-languages}.


\paragraph*{Continuous Extensions}
Lochbihler and H{\"{o}}lzl~\cite{DBLP:conf/itp/LochbihlerH14} employ
concepts from domain theory and topology to define the filter
operation on streams as a continuous extension (the ``consumer view'')
of a function on finite lists, and reduce proofs about it to induction
over lists. AlgCo does not explicitly require any concepts from
topology and generalizes from stream transformers to a wide variety of
continuous mappings over coinductive structures including real-valued
semantics (Section~\ref{subsec:discrete-distribution-samplers}) and
propositional functions
(Section~\ref{subsec:cocontinuous-properties}).

Rusu and Nowak take a similar approach for defining corecursive
functions based on finite approximations of CPO
elements~\cite{DBLP:conf/ecoop/RusuN22} to define $\mathsf{filter}$
and $\mathsf{mirror}$ operations on streams and rose trees,
respectively. We identify the underlying abstraction of algebraic CPO
and provide a principled and coherent story around the composition,
verification, and execution of such functions within the AlgCo
framework.

\section{Conclusion}
\label{sec:conclusion}

This paper presents AlgCo (Algebraic Coinductives), a formal
type-theoretic framework for reasoning about continuous functions over
the class of coinductive structures that form algebraic CPOs. We
introduce the basic concepts of the framework and provide a Coq
implementation in the \href{https://github.com/bagnalla/algco}{AlgCo}
library. We demonstrate its usefulness by implementing and verifying a
number of illustrative examples including a stream variant of the
sieve of Eratosthenes, a regular expression framework based on
Brzozowski derivatives, and expected value reasoning over discrete
distribution samplers in the random bit model.

\clearpage

\bibliographystyle{IEEEtran}


\bibliography{main}

\appendix 

\section{Algebraic CPO Proof Principles}
\label{app:algebraic-cpo-proof-principles}

The primary \textit{definition} principle for continuous extensions:

\begin{theorem}[\link{aCPO}{\cwpcozincl}{Existence of continuous extensions}]
  \label{theorem:exists-continuous-extension}
  
  Let $A$ be an algebraic CPO, $C$ any CPO, and $f
  : \basis{A} \rightarrow C$ a monotone function. Define
  \[ \link{aCPO}{\cwpco}{\co{f}} \:
  a \triangleq \sup{(f \circ \mathsf{idl} \: a)}. \]
  Then,
  \begin{equation}
    \label{eq:exists-continuous-extension}
    \co{f} \circ \mathsf{incl} = f.
  \end{equation}

  I.e., the following diagram commutes:
  
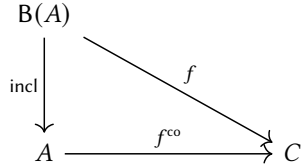
\[
  \begin{tikzcd}
    \basis{A} \arrow[rrrdd, "f"] \arrow[dd, "\mathsf{incl}"'] &  &  &   \\
    &  &  &   \\
    A \arrow[rrr, "\co{f}"]                                              &  &  & C
  \end{tikzcd}
  \]
\end{theorem}


The uniqueness of continuous extensions yields the
fundamental \textit{proof} principle:

\begin{theorem}[\link{aCPO}{\cwpcozunique}{Uniqueness of continuous extensions}]
  \label{theorem:unique-continuous-extension}
  
  Let $A$ be an algebraic CPO, $C$ any CPO, $f : \basis{A} \rightarrow
  C$ a monotone function on the basis of $A$, and $g : A \rightarrow
  C$ a continuous function on $A$ such that:
  \[ g \circ \mathsf{incl} = f, \]
  
  i.e., that the following diagram commutes:
  
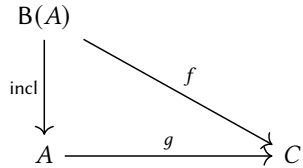
\[
  \begin{tikzcd}
    \basis{A} \arrow[rrrdd, "f"] \arrow[dd, "\mathsf{incl}"'] &  &  &   \\
    &  &  &   \\
    A \arrow[rrr, "g"]                                              &  &  & C
  \end{tikzcd}
  \]
  
  Then, $g = \co{f}$.
\end{theorem}

The following generalized point-wise form of the equivalence principle
is more powerful than
Lemma~\ref{lemma:equivalence-continuous-extension} because it
remembers that the basis elements in the antecedent are approximations
of the CPO elements appearing in the goal, which may be important when
the equality being proved is contingent on some precondition.

\begin{lemma}[\link{aCPO}{\cwpProperzco}{Generalized equivalence of continuous extensions}]
  \label{lemma:generalized-equivalence-continuous-extension}
  
  Let $A$ and $B$ be algebraic CPOs, $C$ any CPO, $f
  : \basis{A} \rightarrow C$, $g : \basis{B} \rightarrow C$ monotone
  functions, $a : A$ and $b : B$. Then,
  \[ \forall \: i : \nat, \: f \: (\mathsf{idl} \: a \: i) = g \:
  (\mathsf{idl} \: b \: i) \Rightarrow \co{f} \: a = \co{g} \: b. \]
\end{lemma}

\begin{corollary}[\link{aCPO}{\cwpcontinuouszcozincl}{Representation of continuous functions}]
  \label{corollary:representation}
  
  Let $A$ be an algebraic CPO, $C$ any CPO, and $g : A \rightarrow C$
  a continuous function. Then,
  \[ g = \co{(g \circ \mathsf{incl})}. \]
\end{corollary}

Corollary \ref{corollary:representation}, a fairly direct consequence
of theorem \ref{theorem:unique-continuous-extension}, shows that every
continuous function $g : A \rightarrow C$ can be written as the
continuous extension $\co{(g \circ \mathsf{incl})}$. I.e., every
continuous function $g : A \rightarrow C$ can be expressed as the
continuous extension of a monotone basis function. It follows that any
two continuous functions are equal when their restrictions to the
basis are equal:


\begin{corollary}[\link{aCPO}{\cwpcontinuouszind}{Equivalence of continuous functions}]
  \label{corollary:equivalence-continuous}
  
  Let $A$ be an algebraic CPO with basis $B$, $C$ any CPO, and $f, g :
  A \rightarrow C$ continuous functions. Then,
  \[ f \circ \mathsf{incl} = g \circ \mathsf{incl} \Rightarrow f =
  g \]
\end{corollary}

\section{Coinductive Extensionality}
\label{app:coinductive-extensionality}

The usual notion of propositional (Leibniz) equality in Coq is too
weak to prove equalities over coinductive types such as $\conat$
(Definition~\ref{def:conat}). For example, suppose we define a
function $\mathsf{coplus}
: \conat \rightarrow \conat \rightarrow \conat$ for taking the sum of
two conats. We quickly become stuck when trying to prove basic
properties such as commutativity: $\forall n \:
m, \: \mathsf{coplus} \: n \: m = \mathsf{coplus} \: m \:
n$. Typically the proof would proceed by induction on either $n$ or
$m$, but here neither term is inductive. We cannot use coinduction either
because the goal is not coinductive.

An alternative is to define a coinductive bisimulation relation that holds
between $n$ and $m$ iff they are structurally identical:

\begin{definition}[\link{conat}{\cwpconatzeq}{$\conat$ Equivalence}]
  \label{def:conat-eq}
  
  Define $=_{\conat} : \rel{\conat}{\conat}$ coinductively by the
  inference rules:
  \begin{mathpar}
    \inferrule [\link{conat}{\cwpconatzeqzzero}{$=_{\conat}$-zero}]
               {  }
               { \cons{cozero}{} =_{\conat} \cons{cozero}{} }
    \and
    \inferrule [\link{conat}{\cwpconatzeqzsucc}{$=_{\conat}$-succ}]
               { n =_{\conat} m }
               { \cons{cosucc} \: n =_{\conat} \cons{cosucc} \: m }
  \end{mathpar}
\end{definition}


Now we can prove commutativity up to $=_{\conat}$ via coinduction:
$\forall n \: m, \: \mathsf{coplus} \: n \: m
=_{\conat} \mathsf{coplus} \: m \: n$. The problem with this approach,
however, is that if we want to rewrite by such equations then we have
to explicitly prove that all of our operations on $\conat$ are proper
wrt.~$=_{\conat}$. But note that $=_{\conat}$ is carefully designed to
coincide exactly with Leibniz equality. Although it cannot be proved
within Coq, we can assert this fact in the form of an extensionality
axiom (Axiom~\ref{axiom:conat-extensionality}) that deduces Leibniz
equality on conats from proofs of bisimilarity (and thus easily
rewrite by them without the need for any $\mathsf{Proper}$ instances).

To gain confidence in the soundness of
Axiom~\ref{axiom:conat-extensionality}, notice that $\conat$ modulo
$=_{\conat}$ is isomorphic to the type $\nat + 1$,
because \link{conat}{\cwpconatzfinitezorzomega}{every conat is either
a finite natural number or the infinite conat
$\omega_\nat$}. Let \link{conat}{\cwpconatztozconatx}{$\mathsf{sect}
: \conat \rightarrow \nat + 1$}
and \link{conat}{\cwpconatxztozconat}{$\mathsf{retr} : \nat +
1 \rightarrow \conat$} witness this isomorphism. We
may \link{conat}{\cwpconatzext}{derive conat extensionality as a
theorem} from \link{conat}{\cwpconatzeqzaxiom}{one side of the
isomorphism:} $\forall n : \conat, \: \mathsf{retr} \:
(\mathsf{sect} \: n) = n$. That is, conat extensionality may be
derived from the fact that injecting a conat into $\nat + 1$ and then
projecting it back reproduces the original conat.

Similar arguments can be made for streams, cotries, and
cotrees. See~\cite[Section 2.2.2]{DBLP:phd/hal/Boulier18}
and~\cite{gross2023extensionality} for more discussion on soundness of
extensionality axioms for coinductive types.

\section{Relating Cotrees to Interaction Trees}
\label{app:relating-cotrees-to-itrees}

The \zar system~\cite{https://doi.org/10.48550/arxiv.2211.06747} does
not extract cotrees directly for execution. Rather, it compiles
probabilistic programs to interaction trees~\cite{xia2019interaction}
(itrees) for sampling execution in OCaml and relates them to
equivalent cotree constructions from which correctness guarantees
(e.g., Theorem~\ref{theorem:equidistribution}) obtained via AlgCo are
transported. Interaction trees contain $\mathsf{Tau}$ nodes (as
described in the introduction of this paper), guaranteeing
productivity at the cost the possibility of infinite $\mathsf{Tau}$
sequences (appearing anywhere $\bot_{\cotree{A}}$ would have appeared
in the cotree representation). \zar thus includes $\mathsf{Tau}$ nodes
in the definition of cotrees for ease of relating them with
corresponding itrees.



The main challenge in relating the two representations is to provide
an iteration construct for cotrees to match the $\mathsf{ITree.iter}$
combinator of the itree library:

\begin{definition}[\link{cotree}{\cwpcotreeziter}{$\mathsf{iter_{\cotree{A}}}$}]
  \label{def:cotree-iter}

  For types $I$ and $A$, $f : I \rightarrow \cotree{I + A}$, and $z :
  I$, define $\mathsf{iter_{\cotree{A}}} \: f \: z
  : \cotree{A} \triangleq \sup{(\mathsf{F}^n \:
  (\lambda \_. \: \cons{cobot})) \: z}$, where $\mathsf{F} :
  (I \rightarrow \cotree{A}) \rightarrow I \rightarrow \cotree{A}$ is
  given by:
  \begin{align*}
    & \mathsf{F} \: g \: i \triangleq f \: i \bind \lambda
    lr. \: \text{match } lr \text{ with } \\
    & \hspace{80pt} \mid \mathsf{inl} \: j \Rightarrow g \: j \\
    & \hspace{80pt} \mid \mathsf{inr} \:
    x \Rightarrow \cons{coleaf} \: x \\
    & \hspace{83pt} \text{end.}
  \end{align*}
\end{definition}

Figure~\ref{fig:iter-haskell} shows a Haskell extraction primitive for
$\mathsf{iter_{\cotree{A}}}$, justified by the following computation
rule:

\begin{lemma}[\link{cotree}{\cwpcotreeziterzunfold}{$\mathsf{iter_{\cotree{A}}}$ computation rule}]
  \label{def:iter-computation}

  Let $I$ and $A$ be types, $f : I \rightarrow \cotree{I + A}$, and $i
  : I$. Then,
  \begin{align*}
    & \mathsf{iter_{\cotree{A}}} \: f \: i = f \: i \bind \lambda
    lr. \: \text{match } lr \text{ with } \\
    & \hspace{100pt} \mid \mathsf{inl} \:
    j \Rightarrow \mathsf{iter_{\cotree{A}}} \: f \: j \\
    & \hspace{100pt} \mid \mathsf{inr} \:
    x \Rightarrow \cons{coleaf} \: x \\
    & \hspace{103pt} \text{end.}
  \end{align*}
\end{lemma}

Further detail on relating itrees and cotrees can be found in the
sources of the \zar system.

\begin{figure}
  \centering
  \begin{tabular}{c}
\begin{lstlisting}
cotree_iter =
  \ f i ->
    cotree_bind (f i) (\ lr ->
      case lr of
        Inl j -> cotree_iter f j
        Inr x -> coleaf x)
\end{lstlisting}
  \end{tabular}
  \caption{Haskell extraction primitive for
  $\mathsf{iter_{\cotree{A}}}$.}
  \label{fig:iter-haskell}
\end{figure}

\end{document}